\documentclass[aps,prb,twocolumn,showpacs,superscriptaddress,longbibliography]{revtex4-2}
\usepackage{amsfonts}
\usepackage{amsmath}
\usepackage{graphicx}
\usepackage{bm}
\usepackage{soul}
\usepackage{ulem}
\usepackage{amssymb}
\usepackage{dcolumn}
\usepackage{color}
\usepackage{multirow}
\usepackage{booktabs}
\usepackage[colorlinks,
linkcolor=blue,
anchorcolor=blue,
citecolor=blue,
urlcolor=blue,
]{hyperref}
\begin{document}

\title{Anomalous Hall and Nernst effects driven by static and fluctuating spin chiralities on Kagome lattice}

\author{He-Lin Li}
\affiliation{School of Physical Sciences, University of Chinese Academy of Sciences, Beijing 100049, China}

\author{Zhen-Gang Zhu}
\email{zgzhu@ucas.ac.cn}
\affiliation{School of Physical Sciences, University of Chinese Academy of Sciences, Beijing 100049, China}
\affiliation{School of Electronic, Electrical and Communication Engineering, University of Chinese Academy of Sciences, Beijing 100049, China}

\author{Gang Su}
\email{gsu@ucas.ac.cn}
\affiliation{Institute of Theoretical Physics, Chinese Academy of Sciences, Beijing 100190, China}
\affiliation{Kavli Institute for Theoretical Sciences, University of Chinese Academy of Sciences, Beijing 100190, China}

\begin{abstract}
We theoretically investigate the anomalous Hall and Nernst effects (AHE and ANE) in a two dimensional Kagome lattice to uncover the distinct roles of static and fluctuating scalar spin chiralities.
Employing Monte Carlo simulations incorporating with a tight binding Hamiltonian via the $s\text{-}d$ exchange interaction, we explicitly evaluate the anomalous transport coefficients.
A key finding is the systematic disentanglement of the macroscopic responses into an intrinsic contribution, governed by momentum space Berry curvature induced by static chirality, and an extrinsic skew scattering contribution driven by real space dynamical spin fluctuations.
We demonstrate a pronounced mechanistic crossover: deep in magnetically ordered phases like the skyrmion crystal, the intrinsic Berry curvature dictates the transport behavior.
However, approaching the magnetic order-disorder critical regime, strong thermal fluctuations disrupt static noncoplanar spin configurations, drastically suppressing intrinsic responses.
Here, dynamical chiral fluctuations emerge as the dominant driving force.
By delineating the phase regimes governed by static versus fluctuating chiralities, this work elucidates the distinct microscopic mechanisms dictating anomalous transport in frustrated magnetic systems.


\end{abstract}
\maketitle{}

\section{Introduction}

Geometrically frustrated structures, with the Kagome lattice as a prominent example, have recently become ideal platforms for investigating topological quantum phenomena, owing to their distinctive atomic arrangements and strong electronic correlations \cite{Balents_2010,Yin_2022}.
The Kagome band structure hosts intrinsic flat bands, Dirac cones, and van Hove singularities—key features that enable the emergence of a rich variety of topological phases stabilized by magnetic coupling. 
As macroscopic signatures of such non-trivial band topology, anomalous transport responses, including the anomalous Hall effect (AHE) in charge transport and the anomalous Nernst effect (ANE) in thermoelectric transport, have attracted tremendous research attention. These effects hold great promise for next-generation low-power spintronic components and thermal energy harvesting devices \cite{Nagaosa_2010,Sakai_2018,Chen_2022,Mizuguchi_2019}.


Conventionally, anomalous transport phenomena in magnetic materials are understood to originate primarily from spin-orbit coupling (SOC) \cite{Nagaosa_2010,Xiao_2010}.
However, in noncoplanar magnetic systems lacking SOC, nontrivial real space spin configurations, including skyrmions and noncoplanar antiferromagnetic states, can host a finite scalar spin chirality defined as $\chi_{ijk} = \boldsymbol{S}_i \cdot (\boldsymbol{S}_j \times \boldsymbol{S}_k)$.
This scalar spin chirality acts as an effective magnetic field that imparts a nontrivial Berry phase to itinerant electrons, thereby driving the topological Hall effect (THE) and its thermoelectric counterpart, the topological Nernst effect (TNE) \cite{Tatara_2002, Nagaosa_2003, Bruno_2004, Nakazawa_2018, Ueda_2012, Neubauer_2009, Metalidis_2006, Verma_2022,Go_2025,Zhang_2021,Oike_2022,Li_2019,Kolincio_2021,Vistoli_2018,Nakazawa_2014,Onoda_2004,Vir_2019,Feng_2020}.
From a general theoretical viewpoint, the full physical origins underlying the AHE and ANE fall into two categories: an intrinsic channel dictated by momentum-space Berry curvature, and extrinsic scattering contributions—primarily skew scattering and side-jump scattering—arising from impurities or dynamic spin fluctuations \cite{Nagaosa_2010,Xiao_2010}.


Despite these theoretical advances, capturing both the static and dynamical manifestations of scalar spin chirality within a unified framework remains a formidable challenge in the field of topological magnetism.
Specifically, it is notoriously difficult to quantitatively disentangle the intrinsic transport governed by momentum space Berry curvature from the extrinsic scattering driven by real space spin fluctuations.


To address these theoretical challenges, we systematically investigate the anomalous Hall and Nernst conductivities (AHC and ANC) in a two dimensional Kagome lattice coupled via the $s\text{-}d$ exchange interaction.
Employing Monte Carlo (MC) simulations combined with a tight binding model, we precisely determine the thermal expectation values of the local spins and explicitly extract the thermal fluctuations of the scalar spin chirality under varying temperatures and magnetic fields.
By incorporating the static spin averages into the tight binding Hamiltonian, we calculate the transport coefficients governed by the intrinsic momentum space Berry curvature.
Meanwhile, based on analytical transport equations and the extracted chiral fluctuations, we separately calculate extrinsic transport contributions generated by chiral skew scattering.


Our results reveal a pronounced mechanistic crossover governing the anomalous transport across different magnetic regimes.
Deep within the topologically nontrivial skyrmion phase, the system exhibits a robust static noncoplanar spin configuration.
This static scalar chirality generates a nontrivial Berry curvature in momentum space, which fundamentally drives the intrinsic AHC and ANC.
Conversely, as the system approaches the magnetic order-disorder phase boundary, intense thermal fluctuations disrupt this static noncoplanar order, leading to a severe suppression of the intrinsic transport coefficients.
Remarkably, the macroscopic anomalous transport persists in this critical region, predominantly sustained by the extrinsic skew scattering mechanism arising from emergent thermal chiral fluctuations.
In a word, we clarify the competing relationship between intrinsic and extrinsic mechanisms across various phase regimes, offering a novel theoretical perspective for understanding critical topological transport in frustrated magnets.


\section{Kondo lattice model}

To describe the exchange coupling between the localized spin moments $\boldsymbol{S}_i$ ($|\boldsymbol{S}_i|=1$) and the itinerant electrons, we employ the canonical Kondo lattice, or $s\text{-}d$ exchange, where the Hamiltonian \cite{Onoda_2004, Tatara_2002} reads $H = H_{0}+H_{\mathrm{K}}$, and
\begin{eqnarray}
H_{0} &=& \sum_{\langle ij\rangle} t_{ij} c_{i\alpha}^{\dagger} c_{j\alpha}, \notag\\
H_{\mathrm{K}} &=& J_{\mathrm{K}} \sum_{i} c_{i\alpha}^{\dagger} \boldsymbol{\sigma}_{\alpha\beta} c_{i\beta} \cdot \boldsymbol{S}_{i},
\label{Kondo_Hamilton}
\end{eqnarray}
where $c_{i\alpha}^{\dagger}$ ($c_{i\alpha}$) represents the creation (annihilation) operator of an itinerant electron on site $i$ with spin $\alpha$.
The first term characterizes the kinetic energy of the itinerant electrons, where $t_{ij}$ represents the hopping transfer integral between nearest neighbor sites $\langle ij \rangle$.
The second term denotes the local exchange interaction between the conduction electrons and the localized magnetic moments.
Here, $J_{\mathrm{K}}$ is the effective exchange coupling strength, $\boldsymbol{\sigma}$ is the vector of Pauli matrices, and $\boldsymbol{S}_i$ is the classical localized spin vector.
Note that the repeated spin indices $\alpha$ and $\beta$ implicitly denote summation.

\subsection{Intrinsic AHC and ANC}
To evaluate the intrinsic anomalous transport properties, we consider the static limit where the itinerant electrons couple to the thermal expectation values of the localized spin moments, $\langle \boldsymbol{S}_i \rangle$, obtained from the classical Monte Carlo simulations of the spin lattice (introduced in Sec. IV).
By substituting these static spin configurations into the tight binding Hamiltonian and performing a Fourier transform, we obtain the momentum space Hamiltonian $H(\boldsymbol{k})$.
Diagonalizing $H(\boldsymbol{k})$ yields the energy eigenvalues $\varepsilon_{n\boldsymbol{k}}$ and the corresponding Bloch eigenstates $|n\boldsymbol{k}\rangle$ for the $n$-th band.

The intrinsic AHC is fundamentally determined by the geometric properties of the electronic wavefunctions, which are quantified by the momentum space Berry curvature \cite{Nagaosa_2010, Xiao_2010}.
According to linear response theory (the Kubo formula), the intrinsic AHC $\sigma_{xy}^{\mathrm{int}}$ can be expressed as an integral over the Brillouin zone:
\begin{equation}
\sigma_{xy}^{\mathrm{int}} = -\frac{e^2}{\hbar} \int \frac{d^3k}{(2\pi)^3} \sum_{n} f(\varepsilon_{n\boldsymbol{k}}) \Omega^z_n(\boldsymbol{k}), \label{AHC_int}
\end{equation}
where $e$ is the elementary charge, $\hbar$ is the reduced Planck constant, $f(\varepsilon_{n\boldsymbol{k}}) = 1 / \{\exp{[(\varepsilon_{n\boldsymbol{k}} - \mu)/k_{\mathrm{B}} T]} + 1\}$ is the Fermi Dirac distribution function with chemical potential $\mu$ and temperature $T$, and $\Omega^z_n(\boldsymbol{k})$ is the $z$ component of the momentum space Berry curvature for the $n$-th band.
This $z$ component is explicitly calculated as:
\begin{equation}
\Omega^z_n(\boldsymbol{k}) = -2 \mathrm{Im} \sum_{m \neq n} \frac{\langle n\boldsymbol{k}|v_x|m\boldsymbol{k}\rangle \langle m\boldsymbol{k}|v_y|n\boldsymbol{k}\rangle}{(\varepsilon_{n\boldsymbol{k}} - \varepsilon_{m\boldsymbol{k}})^2},
\end{equation}
where $v_{x,y} = \frac{\partial H(\boldsymbol{k})}{\partial k_{x,y}}$ denote the velocity matrix elements along the $x$ and $y$ directions.


Furthermore, the intrinsic ANC, which characterizes the transverse thermoelectric response induced by a longitudinal temperature gradient, can be evaluated within the generalized Berry phase formalism \cite{Xiao_2006_PRL, Xiao_2010}.
The transverse thermoelectric conductivity $\alpha_{xy}^{\mathrm{int}}$ is given by the integral of the Berry curvature weighted by the entropy density of the itinerant electrons:
\begin{equation}
  \alpha_{xy}^{\mathrm{int}} = \frac{e k_{\mathrm{B}}}{\hbar} \int \frac{d^3k}{(2\pi)^3} \sum_{n} s(\varepsilon_{n\boldsymbol{k}}) \Omega^z_n(\boldsymbol{k}), \label{ANE_int}
\end{equation}
where $k_{\mathrm{B}}$ is the Boltzmann constant, and $s(\varepsilon_{n\boldsymbol{k}})$ is the entropy density function for the $n$-th band, defined as:
\begin{align}
s(\varepsilon_{n\boldsymbol{k}}) =\, & -f(\varepsilon_{n\boldsymbol{k}}) \ln f(\varepsilon_{n\boldsymbol{k}}) \nonumber\\
 - &  [1 - f(\varepsilon_{n\boldsymbol{k}})] \ln [1 - f(\varepsilon_{n\boldsymbol{k}})].\label{5}
\end{align}
%
These equations establish the theoretical framework for the intrinsic mechanism in our model, where the noncoplanar real space spin configurations induce a nontrivial momentum space Berry curvature, 
which governs the intrinsic AHC and ANC, manifesting as a topological response that persists even in the absence of atomic spin orbit coupling.  

\subsection{Extrinsic contributions from chiral spin fluctuations}
Beyond the intrinsic anomalous transport driven by static spin configurations (leading to scalar spin chirality), itinerant electrons also undergo asymmetric scattering induced by the thermal fluctuations of localized spins.
Following the theoretical framework established by Ishizuka and Nagaosa \cite{Ishizuka_2018}, the thermal fluctuations of scalar spin chirality in noncoplanar magnets and strongly frustrated systems act as a novel topological scattering source,
and give rise to a substantial extrinsic anomalous Hall effect predominantly governed by the skew scattering mechanism, which persists remarkably even in the absence of spin-orbit coupling.
For brevity, we call the thermal fluctuations of scalar spin chirality as dynamical spin chirality (DSC).

To microscopically formulate this DSC-induced skew scattering, we return to the Kondo lattice Hamiltonian.
By treating the $s\text{-}d$ exchange interaction as a perturbation, we derive the scattering transition rate of the itinerant electrons.
Under the quasistatic approximation, where the temporal variations of the thermal spin fluctuations are much slower than the motion of itinerant electrons, the scattering process is considered entirely elastic.
According to Fermi's golden rule, the transition probability per unit time for an electron scattering from an initial state $|\boldsymbol{k}\alpha\rangle$ to a final state $|\boldsymbol{k}^{\prime}\beta\rangle$ is expressed as:
\begin{equation}
  W_{\boldsymbol{k}\alpha\to\boldsymbol{k}^{\prime}\beta} = \frac{2\pi}{\hbar} \left|F_{\beta\alpha}(\boldsymbol{k}^{\prime},\boldsymbol{k})\right|^2 \delta(\varepsilon_{\boldsymbol{k}\alpha} - \varepsilon_{\boldsymbol{k}^{\prime}\beta}),
\end{equation}
where $F_{\beta\alpha}(\boldsymbol{k}^{\prime},\boldsymbol{k})$ represents the scattering amplitude between the initial and final states.
The Dirac delta function $\delta(\varepsilon_{\boldsymbol{k}\alpha} - \varepsilon_{\boldsymbol{k}^{\prime}\beta})$ rigorously enforces energy conservation during this elastic scattering process.

To capture the asymmetric scattering mechanism responsible for the extrinsic anomalous Hall effect, the first Born approximation is strictly insufficient because it yields a symmetric transition rate.
Therefore, it is essential to expand the scattering amplitude $F_{\beta\alpha}(\boldsymbol{k}^{\prime},\boldsymbol{k})$ up to the second Born approximation:
\begin{equation} F_{\beta\alpha}(\boldsymbol{k}^{\prime},\boldsymbol{k}) = F_{\beta\alpha}^{(1)}(\boldsymbol{k}^{\prime},\boldsymbol{k}) + F_{\beta\alpha}^{(2)}(\boldsymbol{k}^{\prime},\boldsymbol{k}), \end{equation}
where the first order and second order scattering matrix elements are given, respectively, by:
\begin{align} F_{\beta\alpha}^{(1)}(\boldsymbol{k}^{\prime},\boldsymbol{k}) &= \langle\boldsymbol{k}^{\prime}\beta|H_{\mathrm{K}}|\boldsymbol{k}\alpha\rangle, \\
  F_{\beta\alpha}^{(2)}(\boldsymbol{k}^{\prime},\boldsymbol{k}) &= \langle\boldsymbol{k}^{\prime}\beta|H_{\mathrm{K}} G(\varepsilon_{\boldsymbol{k}}) H_{\mathrm{K}}|\boldsymbol{k}\alpha\rangle. \end{align}
In these expressions, $H_{\mathrm{K}}$ is treated as a perturbation.
Furthermore, the term $G(\varepsilon_{\boldsymbol{k}}) = \lim_{\eta \to 0^+} (\varepsilon_{\boldsymbol{k}} - H_0 + i\eta)^{-1}$ denotes the unperturbed retarded Green's function of the itinerant electrons at the initial state energy $\varepsilon_{\boldsymbol{k}}$. 

To explicitly quantify the skew scattering process, we extract the antisymmetric component of the transition probability.
Within the low energy effective mass approximation, where the itinerant electrons are described by a parabolic dispersion $\varepsilon_{\boldsymbol{k}} = \hbar^2 k^2 / 2m$, we evaluate the breaking of detailed balance between the forward and backward scattering processes.
Consequently, the antisymmetric scattering rate is defined as:
\begin{equation}
  2w_{\boldsymbol{k}\alpha\to\boldsymbol{k}^{\prime}\beta}^{-} = W_{\boldsymbol{k}\alpha\to\boldsymbol{k}^{\prime}\beta} - W_{\boldsymbol{k}^{\prime}\beta\to\boldsymbol{k}\alpha}.
\label{Def-wminus}
\end{equation}

By substituting the first order and second order scattering amplitudes into Eq. (\ref{Def-wminus}) and performing extensive algebraic simplifications, we derive (see App. \ref{app-A})
\begin{eqnarray}
w^{-}_{\boldsymbol{k}\alpha\to\boldsymbol{k}^{\prime}\beta} &=&  \frac{m J_{\mathrm{K}}^3 V}{\hbar^3 N^3}  \sum_{l,i\neq j} \left[ (\boldsymbol{S}_l \cdot \boldsymbol{\sigma}_{\beta\alpha})((\boldsymbol{S}_i \times \boldsymbol{S}_j) \cdot \boldsymbol{\sigma}_{\alpha\beta}) \right. \notag\\
& \times & \left. \mathcal{I}_{ijl}(\boldsymbol{k}, \boldsymbol{k}^{\prime}) \frac{i e^{-i k r_{ij}}}{2 r_{ij}} + \mathrm{h.c.} \right]\delta(\varepsilon_{\boldsymbol{k}\alpha} - \varepsilon_{\boldsymbol{k}^{\prime}\beta}). \notag
\end{eqnarray}
A spatial interference factor, $\mathcal{I}_{ijl}(\boldsymbol{k}, \boldsymbol{k}^{\prime})$ is introduced and defined as
\begin{align}
\mathcal{I}_{ijl}(\boldsymbol{k}, \boldsymbol{k}^{\prime}) &= \cos\left[\boldsymbol{k}\cdot(\boldsymbol{R}_l-\boldsymbol{R}_j)-\boldsymbol{k}^{\prime}\cdot(\boldsymbol{R}_l-\boldsymbol{R}_i)\right] \nonumber\\
& - \cos\left[\boldsymbol{k}\cdot(\boldsymbol{R}_l-\boldsymbol{R}_i)-\boldsymbol{k}^{\prime}\cdot(\boldsymbol{R}_l-\boldsymbol{R}_j)\right].
\label{12}
\end{align}
In the form of $w^{-}_{\boldsymbol{k}\alpha\to\boldsymbol{k}^{\prime}\beta}$, the spin dependent matrix element $(\boldsymbol{S}_l \cdot \boldsymbol{\sigma}_{\beta\alpha})[(\boldsymbol{S}_i \times \boldsymbol{S}_j) \cdot \boldsymbol{\sigma}_{\alpha\beta}]$ describes the scattering of conduction electrons by the local noncoplanar magnetic moments.
Upon tracing over the conduction electron spin indices, this term directly reduces to the scalar spin chirality $\boldsymbol{S}_l \cdot (\boldsymbol{S}_i \times \boldsymbol{S}_j)$.
The term $\mathcal{I}_{ijl}(\boldsymbol{k}, \boldsymbol{k}^{\prime})$ serves as a spatial interference factor determined by the relative geometric coordinates of the three localized spins within the scattering cluster.
$m$ is the effective mass of the itinerant electrons, $V$ is the total system volume, and $N$ denotes the total number of lattice sites.
The vectors $\boldsymbol{R}_l$, $\boldsymbol{R}_i$, and $\boldsymbol{R}_j$ are coordinates of the localized spins, while $r_{ij} = |\boldsymbol{R}_i - \boldsymbol{R}_j|$ is the distance between site $i$ and site $j$.
$k = |\boldsymbol{k}| = |\boldsymbol{k}^{\prime}|$ is the magnitude of the wave vector on the constant energy surface, as dictated by the elastic scattering condition.


\subsubsection{Anomalous Hall conductivity from skew scattering}
To derive the steady state nonequilibrium distribution function $g_{\boldsymbol{k}}^\alpha$ of the itinerant electrons,
we substitute the antisymmetric scattering rate into the linearized Boltzmann transport equation \cite{Nagaosa_2010, Nakazawa_2014}, and obtain
\begin{equation}
  e \boldsymbol{v}_{\boldsymbol{k},\alpha} \cdot \boldsymbol{E} f_{0}^{\prime}(\varepsilon_{\boldsymbol{k},\alpha}) = \frac{g_{\boldsymbol{k}}^\alpha}{\tau}
   - \sum_{\beta} \frac{V}{(2\pi)^{3}} \int d^3\boldsymbol{k}^{\prime}  w_{\boldsymbol{k}^{\prime}\beta \to \boldsymbol{k}\alpha}^{-} g_{\boldsymbol{k}^{\prime}}^\beta, \label{eq:boltzmann} \end{equation}
where $\boldsymbol{E}$ is the applied electric field, $\boldsymbol{v}_{\boldsymbol{k},\alpha}$ is the band velocity, $f_{0}^{\prime}(\varepsilon_{\boldsymbol{k},\alpha})$ is the energy derivative of the equilibrium Fermi-Dirac distribution function, and $\tau$ denotes the momentum relaxation time. 
To analytically solve this integro-differential Boltzmann equation, we rely on the standard skew scattering formalism \cite{Tatara_2002, Ishizuka_2018} and assume that the antisymmetric scattering rate derived from the DSC can be cast into the following compact form
\begin{equation}
  w_{\boldsymbol{k}^{\prime}\beta \to \boldsymbol{k}\alpha}^{-} = \tilde{\boldsymbol{V}}_{\alpha\beta}(k) \cdot \frac{\boldsymbol{k} \times \boldsymbol{k}^{\prime}}{k^2} \delta(\varepsilon_{\boldsymbol{k}\alpha} - \varepsilon_{\boldsymbol{k}^{\prime}\beta}),
\label{wminus}
\end{equation}
where $\tilde{\boldsymbol{V}}_{\alpha\beta}(k)$ is an introduced auxiliary vector function. 
The cross product term $\boldsymbol{k} \times \boldsymbol{k}^{\prime}$ explicitly captures the spatial asymmetry for the skew scattering.
Because the elastic scattering constraint imposes $k^{\prime} = k$, integrating out the momentum using the Dirac delta function naturally yields the density of states $\rho(k) = m k / (2\pi^2\hbar^2)$.
Substituting Eq. (\ref{wminus}) into Eq.~(\ref{eq:boltzmann}) and integrating over the solid angle, we have 
\begin{equation}
  g_{\boldsymbol{k}}^\alpha = \tau e \boldsymbol{v}_{\boldsymbol{k},\alpha} \cdot \boldsymbol{E} f_0^{\prime}(\varepsilon_{\boldsymbol{k},\alpha}) + \sum_{\beta} V \tau \frac{\rho(k)}{4\pi} \tilde{\boldsymbol{V}}_{\alpha\beta}(k) \cdot \frac{\boldsymbol{k} \times \boldsymbol{P}_{\beta}(k^{\prime})}{k^2}.
\end{equation}
In this equation, we introduce another auxiliary vector $\boldsymbol{P}_{\beta}(k^{\prime})$, which is defined by the angular integration over the nonequilibrium distribution as
\begin{equation}
  \boldsymbol{P}_{\beta}(k^{\prime}) = \int d\phi d\theta \sin\theta \, \boldsymbol{k}^{\prime} g_{\boldsymbol{k}^{\prime}}^\beta |_{\varepsilon_{\boldsymbol{k}\alpha} = \varepsilon_{\boldsymbol{k}^{\prime}\beta}}.
\end{equation}
%
The charge current density $\boldsymbol{j}$ can be derived by 
\begin{equation}
  \boldsymbol{j} = -e \sum_{\alpha} \int \frac{d^3\boldsymbol{k}}{(2\pi)^3} \boldsymbol{v}_{\boldsymbol{k},\alpha} g_{\boldsymbol{k}}^\alpha.
\end{equation}
In the case of transverse response current $j_x$ driven by a longitudinal electric field $E_y$,
the AHC is then derived as $\sigma_{xy} = j_x / E_y$.

In the low energy regime, assuming a parabolic band dispersion ($v_{\mathrm{F}} = \hbar k_{\mathrm{F}} / m$) and utilizing the relation for the carrier density $n_e = k_{\mathrm{F}}^3 / (6\pi^2)$ appropriate for the spin polarized state, the extrinsic AHC can be simplified into a compact form
\begin{equation}
\sigma^{\mathrm{ext}}_{xy} = e^2 V \tau^2 \frac{2 n_e}{3m} \rho(\varepsilon_{\mathrm{F}}) \left[ \tilde{V}_z^{0}(k_{\mathrm{F}}) + \tilde{V}_z^{1}(k_{\mathrm{F}}) \right],
\label{AHC_ext_3d}
\end{equation}
where $n_e$ denotes the carrier number density, $\rho(\varepsilon_{\mathrm{F}})$ is the density of states at the Fermi level. 
$k_{\mathrm{F}}$ and $v_{\mathrm{F}}$ represent the Fermi wave vector and Fermi velocity, respectively.
$\tilde{V}_z^{0}(k_{\mathrm{F}})$ and $\tilde{V}_z^{1}(k_{\mathrm{F}})$ are the values of $\tilde{V}^{0}(k)$ and $\tilde{V}^{1}(k)$ when $\boldsymbol{n}$ is aligned to $z$ direction and $k=k_{\mathrm{F}}$ in Eq. (\ref{spin_a_f}), meaning the $z$ components of the spin dependent auxiliary functions evaluated at Fermi surface.
These terms originate from expanding the auxiliary vector $\tilde{\boldsymbol{V}}_{\alpha\beta}(k)$ in the basis of Pauli matrices as
\begin{equation}
  \tilde{\boldsymbol{V}}_{\alpha\beta}(k) = \left[ \tilde{V}^0(k)\sigma_0^{\alpha\beta} + \tilde{V}^1(k)\sigma_x^{\alpha\beta} + i\tilde{V}^2(k)\sigma_y^{\alpha\beta} \right] \boldsymbol{n},\label{spin_a_f}
\end{equation}
where $\sigma_0$ is the $2 \times 2$ identity matrix, $\sigma_x$ and $\sigma_y$ are the Pauli matrices, and $\boldsymbol{n}$ represents the spatial orientation vector.

\subsubsection{Anomalous Nernst conductivity induced by antisymmetric scattering}
In this section, we investigate the anomalous Nernst conductivity $\alpha_{xy}$ originating from the same antisymmetric scattering processes discussed in the preceding derivation.
The ANE manifests itself as a transverse charge current density $\boldsymbol{j}$ in response to a longitudinal temperature gradient $-\nabla T$.
%
%
The linearized Boltzmann equation is thus expressed as
\begin{eqnarray}
    \boldsymbol{v}_{\boldsymbol{k},\alpha} \cdot \nabla T  \frac{\partial f_0(\varepsilon_{\boldsymbol{k},\alpha})}{\partial T}
    & =&  -\frac{g_{\boldsymbol{k}}^\alpha}{\tau} + \frac{V}{(2\pi)^{3}}\sum_{\beta} \int d^3\boldsymbol{k}^{\prime} \notag \\
    & \times &    w_{\boldsymbol{k}^{\prime}\beta \to \boldsymbol{k}\alpha}^{-} g_{\boldsymbol{k}^{\prime}}^\beta,
\label{linearBoltz-1}
\end{eqnarray}
where $f_0(\varepsilon_{\boldsymbol{k},\alpha})$ is the equilibrium Fermi-Dirac distribution function.

Via analogous derivation to AHC, the Boltzmann equation is derived in this case as 
\begin{align}
  g_{\boldsymbol{k}}^\alpha = &-\tau \boldsymbol{v}_{\boldsymbol{k},\alpha} \cdot \nabla T \frac{\partial f_0(\varepsilon_{\boldsymbol{k},\alpha})}{\partial T} \nonumber\\
  &+ \sum_{\beta} V \tau \frac{\rho(k)}{4\pi} \tilde{\boldsymbol{V}}_{\alpha\beta}(k) \cdot \frac{\boldsymbol{k} \times \boldsymbol{P}_{\beta}(k^{\prime})}{k^2},
\label{21}
\end{align}
where the first term represents the longitudinal carrier diffusion driven by the temperature gradient; while the second term accounts for the transverse skew scattering contribution giving rise to the ANE.

%
Considering a response charge current $j_x$ driven by $-\nabla_y T$, we derive the ANC $\alpha_{xy}$ via the equation $j_x = \alpha_{xy}(-\nabla_y T)$.
Integrating over the momentum space and performing the spherical angular average ($\int d\Omega \, v_x^2 = \frac{4\pi}{3} v_{\boldsymbol{k}}^2$), we obtain 
\begin{equation}
  \alpha^{\mathrm{ext}}_{xy} = -e V \tau^2 \int \frac{\rho(k)}{9\pi^2} k^2 dk \left[ \tilde{V}_z^{0}(k) + \tilde{V}_z^{1}(k) \right] v_{\boldsymbol{k}}^2 \frac{\partial f_0(\varepsilon_{\boldsymbol{k}})}{\partial T}.
\end{equation}

In the low temperature regime ($k_{\mathrm{B}} T \ll \varepsilon_{\mathrm{F}}$), $\frac{\partial f_0}{\partial T}$ is sharply localized around the Fermi level.
By utilizing the identity $\frac{\partial f_0}{\partial T} = -\frac{\varepsilon-\mu}{T} \frac{\partial f_0}{\partial \varepsilon}$ and evaluating the energy integral via the Sommerfeld expansion, the anomalous Nernst conductivity analytically reduces to the Mott relation (see App. \ref{app-C}):
\begin{equation}
  \alpha^{\mathrm{ext}}_{xy} = -\frac{\pi^2 k_{\mathrm{B}}^2 T}{3e} \left. \frac{d\sigma^{\mathrm{ext}}_{xy}(\varepsilon)}{d\varepsilon} \right|_{\varepsilon=\varepsilon_{\mathrm{F}}},\label{mott_r_ext}
\end{equation}
where $\sigma^{\mathrm{ext}}_{xy}(\varepsilon)$ is the energy dependent anomalous Hall conductivity derived in the previous section (Eq.~\eqref{AHC_ext_3d}).

\begin{figure}[tb]
  \centering
    \includegraphics[width=1\columnwidth]{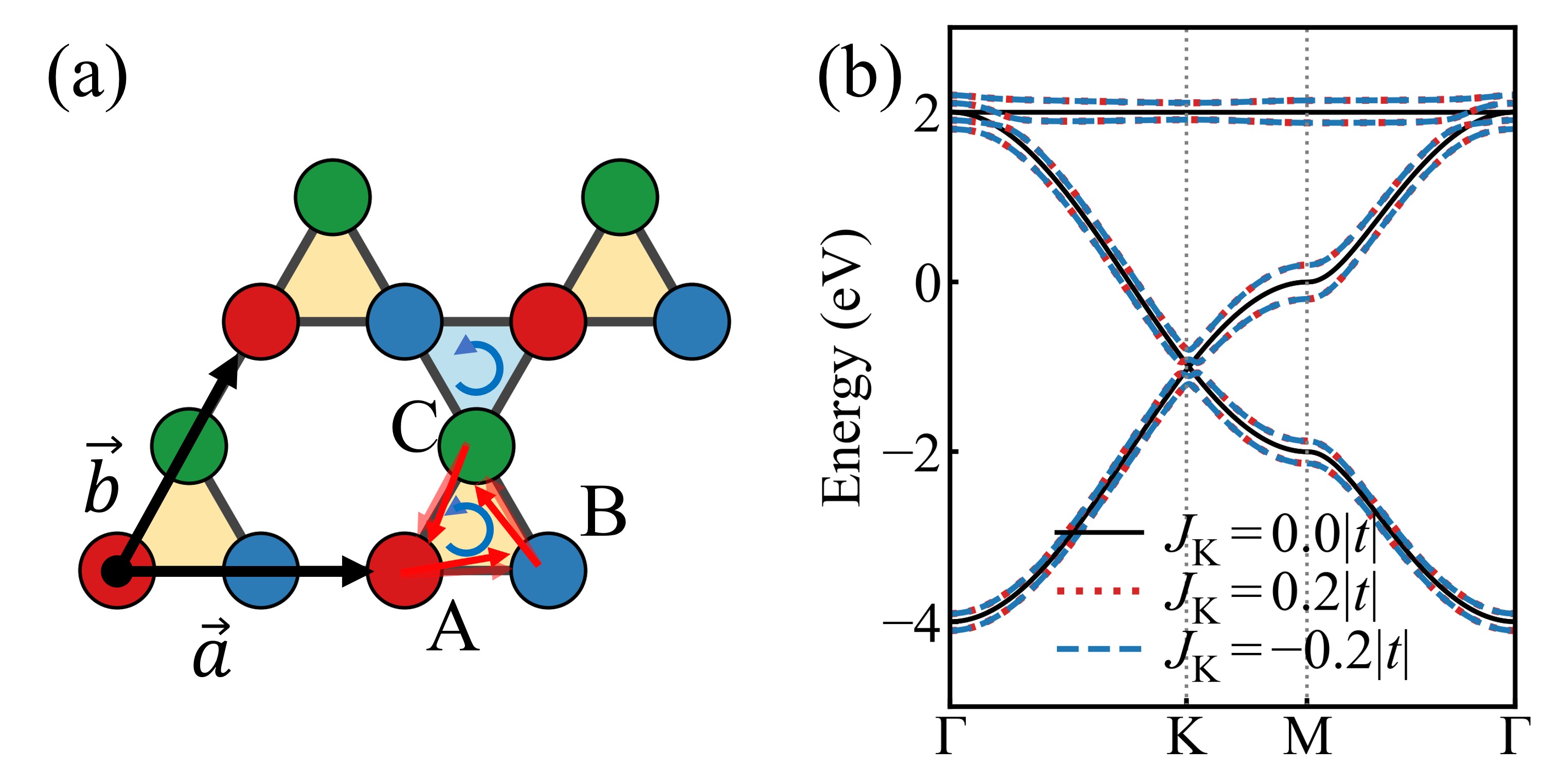}
    \caption{(a) Schematic illustration of the Kagome lattice.
    The black arrows denote the lattice basis vectors $\boldsymbol{a}$ and $\boldsymbol{b}$, while the red, blue, and green spheres represent the three sublattices.
    A representative noncoplanar magnetic configuration with nonzero scalar spin chirality is indicated by red arrows, defined as $\boldsymbol{S}_\mathrm{A} = (\sin\theta, 0, \cos\theta)$, $\boldsymbol{S}_\mathrm{B} = (-\frac{1}{2}\sin\theta, \frac{\sqrt{3}}{2}\sin\theta, \cos\theta)$, and $\boldsymbol{S}_\mathrm{C} = (-\frac{1}{2}\sin\theta, -\frac{\sqrt{3}}{2}\sin\theta, \cos\theta)$ with $\theta=\pi/3$.
    The blue circular arrows specify the counter-clockwise (CCW) orientation used for evaluating the scalar spin chirality.
    (b) Electronic band structure of the Kagome lattice model with a hopping integral $t = -1$~eV.
    The black solid, red dotted, and blue dashed lines represent the $s\text{-}d$ exchange couplings $J_{\mathrm{K}} = 0$, $0.2|t|$, and $-0.2|t|$, respectively. }
  \label{fig_kagome_band}
\end{figure}

\section{Two-dimensional Kagome lattice system}
Having established the general semiclassical framework for anomalous transport, we now apply it to a paradigmatic model, the two-dimensional Kagome lattice \cite{Chen_2014, Ishizuka_2018}.
Due to its inherent strong magnetic frustration \cite{Syromiatnikov_2002, Han_2017}, the Kagome geometry serves as an ideal platform for hosting noncoplanar spin textures and a nonvanishing scalar spin chirality \cite{Ohgushi_2000, Shindou_2001}.

\begin{figure}[tb]
  \centering
    \includegraphics[width=1\columnwidth]{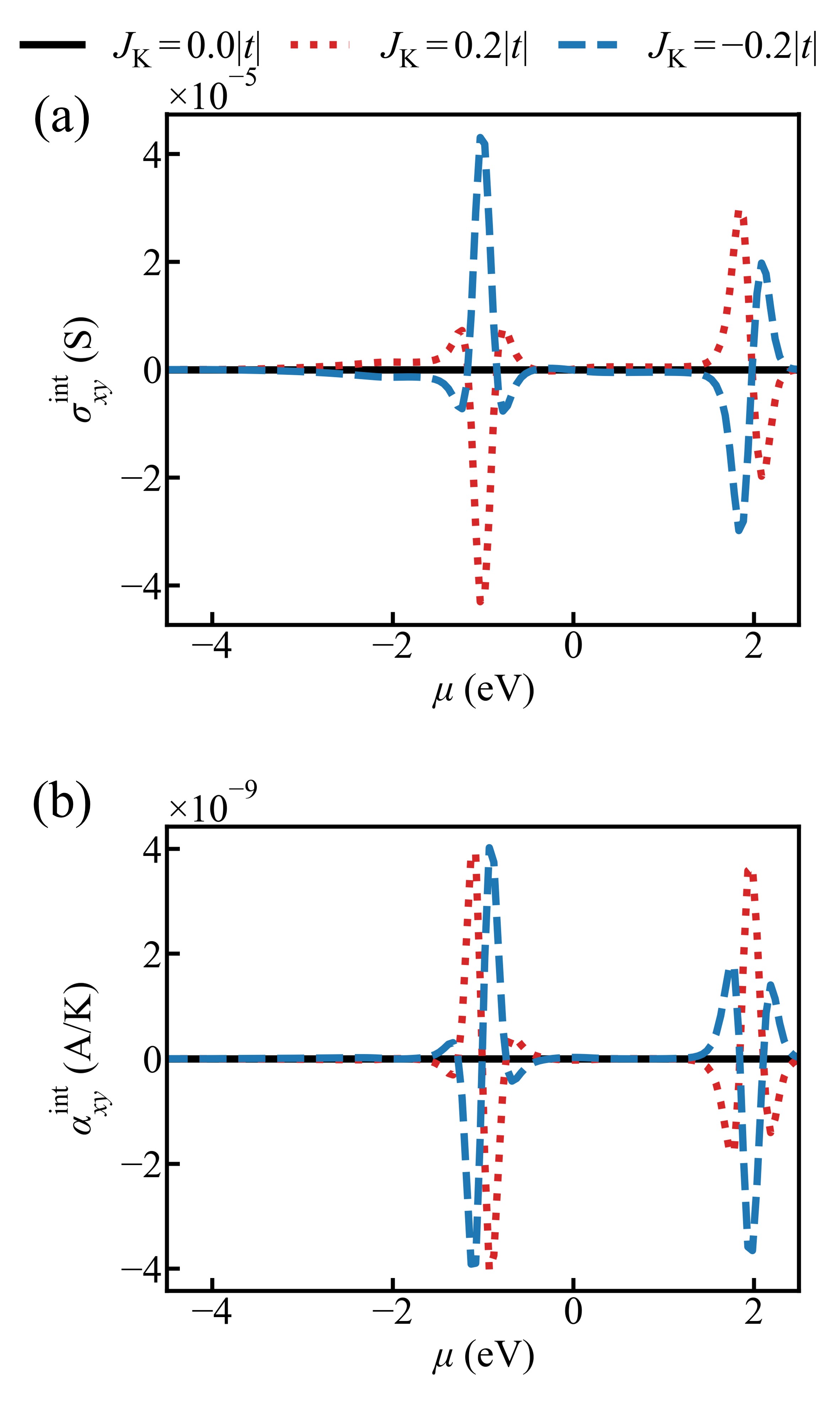}
    \caption{(a) Intrinsic anomalous Hall conductivity $\sigma^{\textrm{int}}_{xy}$ as a function of the chemical potential calculated for various exchange couplings $J_{\mathrm{K}}$ with a hopping integral $t = -1$~eV.
    The black solid, red dotted, and blue dashed lines represent $J_{\mathrm{K}} = 0$, $0.2|t|$, and $-0.2|t|$, respectively.
    (b) Corresponding intrinsic anomalous Nernst conductivity $\alpha^{\textrm{int}}_{xy}$ for the same set of parameters.
    The horizontal axis in both panels denotes the chemical potential.}
  \label{AHC_ANE_diagram}
\end{figure}

Fig.~\ref{fig_kagome_band}(a) is a schematic of the Kagome lattice with sublattices A, B, and C.
Red arrows denote the noncoplanar spin configuration with $120^\circ$ antiferromagnetic order and out of plane canting $\theta = \pi/3$.
Localized spins are $\boldsymbol{S}_{\mathrm{A}} = (\sin\theta, 0, \cos\theta)$, $\boldsymbol{S}_{\mathrm{B}} = (-\frac{1}{2}\sin\theta, \frac{\sqrt{3}}{2}\sin\theta, \cos\theta)$, and $\boldsymbol{S}_{\mathrm{C}} = (-\frac{1}{2}\sin\theta, -\frac{\sqrt{3}}{2}\sin\theta, \cos\theta)$.
Substituting these spins and $t_{ij} = t = -1$ eV into Eq.~\eqref{Kondo_Hamilton} yields the energy bands for various $J_{\mathrm{K}}$ in Fig.~\ref{fig_kagome_band}(b).
Based on these energy bands, intrinsic AHC and ANC in Fig.~\ref{AHC_ANE_diagram} are calculated via Eq.~\eqref{AHC_int} and Eq.~\eqref{ANE_int}.
These results show that noncoplanar spin textures drive intrinsic transport via $s\text{-}d$ exchange without atomic spin orbit coupling.

Besides the intrinsic AHC and ANC (Fig.~\ref{AHC_ANE_diagram}), an extrinsic response exists due to scattering by $H_{\mathrm{K}}$.
The spatial interference factor $\mathcal{I}_{ijl}(\boldsymbol{k}, \boldsymbol{k}^{\prime})$ of the Kagome triangles is expanded via Taylor series in the long wavelength limit ($ka \ll 1$).
Defining relative spatial vectors $\boldsymbol{\delta}_{lj} = \boldsymbol{R}_l - \boldsymbol{R}_j$ and $\boldsymbol{\delta}_{li} = \boldsymbol{R}_l - \boldsymbol{R}_i$ yields the expansion:
\begin{align}
  \mathcal{I}_{ijl}(\boldsymbol{k}, \boldsymbol{k}^{\prime}) &\approx \frac{1}{2} \big[ (\boldsymbol{k}\cdot\boldsymbol{\delta}_{li})^2 - (\boldsymbol{k}\cdot\boldsymbol{\delta}_{lj})^2 \nonumber\\
  & + (\boldsymbol{k}^{\prime}\cdot\boldsymbol{\delta}_{lj})^2 - (\boldsymbol{k}^{\prime}\cdot\boldsymbol{\delta}_{li})^2 \big] \nonumber\\
  & + (\boldsymbol{k}\cdot\boldsymbol{\delta}_{lj})(\boldsymbol{k}^{\prime}\cdot\boldsymbol{\delta}_{li}) - (\boldsymbol{k}\cdot\boldsymbol{\delta}_{li})(\boldsymbol{k}^{\prime}\cdot\boldsymbol{\delta}_{lj}).\label{24}
\end{align}

Applying the identity $(\boldsymbol{A}\cdot\boldsymbol{C})(\boldsymbol{B}\cdot\boldsymbol{D}) - (\boldsymbol{A}\cdot\boldsymbol{D})(\boldsymbol{B}\cdot\boldsymbol{C}) = (\boldsymbol{A}\times\boldsymbol{B})\cdot(\boldsymbol{C}\times\boldsymbol{D})$ to Eq.~\eqref{24} isolates the antisymmetric contribution $(\boldsymbol{k}\times \boldsymbol{k}^{\prime})\cdot(\boldsymbol{\delta}_{lj}\times \boldsymbol{\delta}_{li})$.
For the counterclockwise (CCW) winding indicated by blue circular arrows in Fig.~\ref{fig_kagome_band}(a), this determines the partial scattering rate $w^{\mathrm{CCW}}_{\boldsymbol{k}\alpha\to\boldsymbol{k}^{\prime}\beta}$:
\begin{align}
  w^{\mathrm{CCW}}_{\boldsymbol{k}\alpha\to\boldsymbol{k}^{\prime}\beta} &= \frac{2\pi m J{\mathrm{K}}^3 A}{\hbar^3 N^3} \delta(\varepsilon_{\boldsymbol{k}\alpha} - \varepsilon_{\boldsymbol{k}^{\prime}\beta})  \sum_{(l,i,j) \in \circlearrowleft} \mathrm{Re} \left[ \varGamma_{lij}^{\alpha\beta} \right] \nonumber\\
  & \times \left[ (\boldsymbol{k}\times \boldsymbol{k}^{\prime})\cdot(\boldsymbol{\delta}_{lj}\times \boldsymbol{\delta}_{li}) \right], \label{w_f_2}
\end{align}
where $\sum_{(l,i,j) \in \circlearrowleft}$ denotes the CCW summation over triangular sites, $\varGamma_{lij}^{\alpha\beta} = (\boldsymbol{S}_l\cdot \boldsymbol{\sigma}_{\beta\alpha}) [(\boldsymbol{S}_i\times\boldsymbol{S}_j)\cdot\boldsymbol{\sigma}_{\alpha\beta}]$ is the chirality scattering term \cite{Ishizuka_2018}, and $A$ is the system area.
Clockwise (CW) and CCW scattering contributions are equal as winding reversal introduces compensating minus signs from spatial ($\boldsymbol{\delta}_{jl} \times \boldsymbol{\delta}_{il} = -\boldsymbol{\delta}_{il} \times \boldsymbol{\delta}_{jl}$) and spin ($\boldsymbol{S}_j \times \boldsymbol{S}_i = -\boldsymbol{S}_i \times \boldsymbol{S}_j$) cross products.
Thus, the total antisymmetric scattering rate is $w^{-}_{\boldsymbol{k}\alpha\to\boldsymbol{k}^{\prime}\beta} = 2w^{\mathrm{CCW}}_{\boldsymbol{k}\alpha\to\boldsymbol{k}^{\prime}\beta}$.
Under coordinates $\boldsymbol{R}_A=(0,0,0)$, $\boldsymbol{R}_B=(a,0,0)$, and $\boldsymbol{R}_C=(a/2, \sqrt{3}a/2, 0)$ with lattice constant $a$, $w^{-}_{\boldsymbol{k}\alpha\to\boldsymbol{k}^{\prime}\beta}$ simplifies to:
\begin{align}
  w^{-}_{\boldsymbol{k}\alpha\to\boldsymbol{k}^{\prime}\beta} &= -\frac{2\sqrt{3}\pi m J_{\mathrm{K}}^3 A a^2}{\hbar^3 N^3} \delta(\varepsilon_{\boldsymbol{k}\alpha} - \varepsilon_{\boldsymbol{k}^{\prime}\beta}) (\boldsymbol{k}\times \boldsymbol{k}^{\prime})_z \nonumber\\
  & \times  \sum_{(l,i,j) \in \circlearrowleft} \mathrm{Re} \left[ \varGamma_{lij}^{\alpha\beta} \right].\label{26}
\end{align}

Based on Eq.~\eqref{26}, Eq.~\eqref{spin_a_f} for Kagome triangles simplifies to:
\begin{equation}
  \tilde{V}_z^{0}(k) + \tilde{V}_z^{1}(k) = -\frac{2\pi J_{\mathrm{K}}^3 A}{\hbar^3 N^3} \sqrt{3} a^2 k^2 N \langle \boldsymbol{S}_l \cdot (\boldsymbol{S}_i \times \boldsymbol{S}_j) \rangle,\label{V_eff_2d}
\end{equation}
where $N$ is the number of unit cells.
Substituting Eq.~\eqref{V_eff_2d} into Eq.~\eqref{AHC_ext_3d}, $\sigma^{\mathrm{ext}}_{xy}$ for the two dimensional Kagome lattice simplifies to:
\begin{align}
  \sigma^{\mathrm{ext}}_{xy} &= e^2 A \tau^2 \frac{n_{e,\mathrm{2D}}}{m} \rho_{\mathrm{2D}}(\varepsilon_{\mathrm{F}}) \left[ \tilde{V}_z^{0}(k_{\mathrm{F}}) + \tilde{V}_z^{1}(k_{\mathrm{F}}) \right] \nonumber\\
  &= -\frac{\sqrt{3}e^2\tau^2 A_{\mathrm{cell}}^2 m^3 J_{\mathrm{K}}^3 a^2 \varepsilon_{\mathrm{F}}^2}{\pi \hbar^9} \langle \boldsymbol{S}_l \cdot (\boldsymbol{S}_i \times \boldsymbol{S}_j) \rangle, \label{AHC_ext_2d}
\end{align}
where $A_{\mathrm{cell}}$ is the primitive unit cell area.
In the two dimensional case, $n_{e,\mathrm{2D}} = k_{\mathrm{F}}^2 / (4\pi)$ and $\rho_{\mathrm{2D}}(\varepsilon_{\boldsymbol{k}}) = m / (2\pi\hbar^2)$.
The Mott relation yields the corresponding extrinsic anomalous Nernst conductivity $\alpha^{\mathrm{ext}}_{xy}$ at low temperature:
\begin{eqnarray}
    \alpha^{\mathrm{ext}}_{xy} &=& -\frac{\pi^2 k_{\mathrm{B}}^2 T}{3e} \left. \frac{d\sigma^{\mathrm{ext}}_{xy}(\varepsilon)}{d\varepsilon} \right|_{\varepsilon=\varepsilon_{\mathrm{F}}}  \label{ANE_ext_2d} \\
    &=& \frac{2\pi k_{\mathrm{B}}^2 T}{3e} \frac{\sqrt{3}e^2\tau^2 A_{\mathrm{cell}}^2 m^3 J_{\mathrm{K}}^3 a^2 \varepsilon_{\mathrm{F}}}{\hbar^9} \langle \boldsymbol{S}_l \cdot (\boldsymbol{S}_i \times \boldsymbol{S}_j) \rangle. \notag
\end{eqnarray}

\section{Spin model and fluctuations of spin chirality}
To extract static spin structures and DSC, thermodynamic configurations are generated from the classical spin Hamiltonian on the Kagome lattice
\begin{equation}
  H_{\mathrm{spin}} = -J_{\mathrm{H}} \sum_{\langle i,j \rangle} \boldsymbol{S}_i \cdot \boldsymbol{S}_j + \sum_{\langle i,j \rangle} \boldsymbol{D}_{ij} \cdot (\boldsymbol{S}_i \times \boldsymbol{S}_j) - \boldsymbol{B} \cdot \sum_i \boldsymbol{S}_i, \label{H_spin}
\end{equation}
where $J_{\mathrm{H}}$ is the nearest neighbor Heisenberg exchange.
The second term is the Dzyaloshinskii-Moriya (DM) interaction, stabilizing noncoplanar spin textures of low dimensional systems \cite{Yang_2015, Liang_2020, Koshibae_2016, Hartl_2024}.

To resolve emergent spin phases in the frustrated Kagome geometry, such as skyrmion crystals (SkX) and helical phases \cite{Rosales_2023, Albarracin_2024_ML}, classical Monte Carlo (MC) simulations are employed.
Simulations utilize a lattice with linear dimension $L = 48$ and $N = 3 \times L^2$ sites under periodic boundary conditions. A hybrid update scheme combines Metropolis single spin updates with microcanonical over relaxation steps.
Since the purpose of the simulation is to generate thermodynamic equilibrium configurations of localized magnetic moments, the spins are treated as classical vectors rather than quantum operators. This classical treatment is appropriate for extracting static spin textures and scalar chirality configurations used in the subsequent electronic-structure and transport calculations.
Thermodynamic equilibrium is reached via simulated annealing. For each temperature and magnetic field, $10^5$ sweeps were used for thermalization, followed by $4 \times 10^5$ sweeps for data collection.

Macroscopic transport coefficients are evaluated by extracting two distinct components from the simulated spin configurations.
First, substituting expectation values $\langle \boldsymbol{S}_i \rangle$ into the Hamiltonian Eq.~\eqref{Kondo_Hamilton} yields intrinsic topological Hall signatures.
Calculations utilize a $12 \times 12$ magnetic supercell to account for spatial periodicity. Diagonalization yields the momentum space Berry curvature and intrinsic transport coefficients across the $T\text{-}B$ parameter space.
Second, extrinsic scattering contributions are determined from the fluctuating scalar spin chirality.
For each elementary triangular plaquette, the triple product $\boldsymbol{S}_i \cdot (\boldsymbol{S}_j \times \boldsymbol{S}_k)$ is calculated following the CCW orientation in Fig.~\ref{fig_kagome_band}(a).
The DSC is subsequently defined as the deviation from the static background
\begin{equation}
\delta \chi_{ijk} = \langle \boldsymbol{S}_i \cdot (\boldsymbol{S}_j \times \boldsymbol{S}_k) \rangle - \langle \boldsymbol{S}_i \rangle \cdot (\langle \boldsymbol{S}_j \rangle \times \langle \boldsymbol{S}_k \rangle). \label{delta_chi}
\end{equation}
This term captures the interplay between chiral fluctuations and itinerant electrons \cite{Albarracin_2024_PRB}.
Substituting the spatial average $\delta \chi = \frac{1}{N_{\triangle}} \sum_{\langle ijk \rangle} \delta \chi_{ijk}$ for total scalar spin chirality $\langle \boldsymbol{S}_i \cdot (\boldsymbol{S}_j \times \boldsymbol{S}_k) \rangle$ in Eqs.~\eqref{AHC_ext_2d} and \eqref{ANE_ext_2d}, the transport coefficients due to the DSC-induced skew scattering can be calculated.
To suppress statistical noise from finite size sampling, Gaussian smoothing is applied in deriving the transport coefficients.

\begin{figure}[tb]
  \centering
    \includegraphics[width=1\columnwidth]{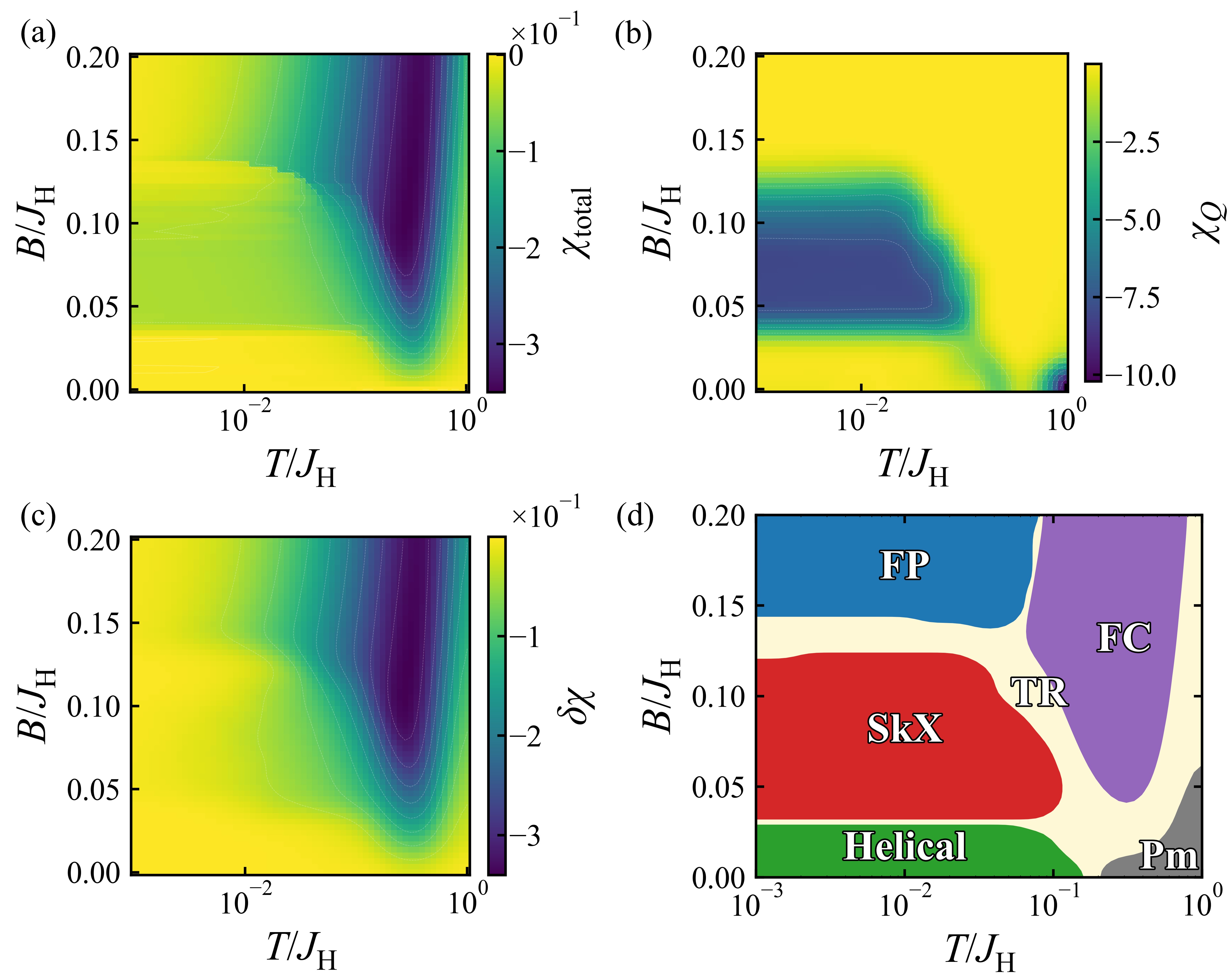}
    \caption{
    (a) Total scalar spin chirality averaged over a single triangle $\chi_{\mathrm{total}} = \langle \boldsymbol{S}_i \cdot (\boldsymbol{S}_j \times \boldsymbol{S}_k) \rangle$ calculated via Monte Carlo (MC) simulations across the $T\text{-}B$ parameter space.
    (b) Macroscopic topological charge $\chi_{Q}$ (static topological background) evaluated using the lattice based solid angle approach.
    (c) Fluctuation driven scalar spin chirality $\delta\chi$, obtained by subtracting the static background from the total chirality, representing the pure thermal fluctuation contribution.
    (d) Comprehensive magnetic phase diagram established from the MC results. The identified regions include: the helical phase, the SkX phase, the field polarized (FP) phase, the paramagnetic (Pm) phase, and the emergent fluctuating chiral (FC) phase. The broad crossover areas between these distinct magnetic states are collectively classified as the transition region (TR).
    }
  \label{phase_diagram}
\end{figure}

\section{Results and Discussion}
To investigate chirality-driven anomalous transport, the thermodynamic phase space of the localized spin model in Eq.~\eqref{H_spin} is calculated.
Monte Carlo results for DM interaction components $D_{xy} = 0.5$ and $D_z = \sqrt{3}$ (in units of $J_{\mathrm{H}}$) are presented in Fig.~\ref{phase_diagram} as a function of temperature and external magnetic field.


Decomposing scalar spin chirality into static and fluctuating parts identifies the microscopic origins of anomalous transport.
Fig.~\ref{phase_diagram}(a) illustrates the total scalar spin chirality $\langle \boldsymbol{S}_i \cdot (\boldsymbol{S}_j \times \boldsymbol{S}_k) \rangle$ across the $T\text{-}B$ parameter space.
This total chirality entangles the static topological background with dynamic thermal spin fluctuations.
To isolate the intrinsic topological contribution, the topological charge $\chi_{Q}$ is calculated using a lattice based approach \cite{Kim_2020}, as shown in Fig.~\ref{phase_diagram}(b).
Following the orientation of the blue arrows in Fig.~\ref{fig_kagome_band}(a), the topological charge is defined as the sum of solid angles $q_{ijk}$ over all elementary triangular plaquettes $\langle ijk \rangle$:
\begin{equation}
  \chi_{Q} = \frac{1}{4\pi}\sum_{\langle ijk\rangle} q_{ijk}.
\end{equation}
For each $\langle ijk \rangle$:
\begin{equation}
  \tan\left(\frac{q_{ijk}}{2}\right) = \frac{\boldsymbol{m}_i\cdot(\boldsymbol{m}_j\times\boldsymbol{m}_k)}{1+\boldsymbol{m}_i\cdot\boldsymbol{m}_j+\boldsymbol{m}_j\cdot\boldsymbol{m}_k+\boldsymbol{m}_k\cdot\boldsymbol{m}_i},
\end{equation}
where $\boldsymbol{m}_i = \langle \boldsymbol{S}_i \rangle / |\langle \boldsymbol{S}_i \rangle|$ is the normalized thermally averaged local spin at site $i$.
This static topological charge quantifies the Berry phase originating from the ordered noncoplanar ground state \cite{Nagaosa_2013}.

Beyond these static topological signatures, subtracting the static background from the total chirality via Eq.~\eqref{delta_chi} isolates the DSC $\delta \chi$, as shown in Fig.~\ref{phase_diagram}(c).
By comprehensively correlating the chiral quantities, real space spin configurations, and macroscopic properties, we construct the magnetic phase diagram shown in Fig.~\ref{phase_diagram}(d).
Quantitatively, the distinct magnetic phases are unambiguously identified using a set of criteria based on the total topological charge $\chi_Q$ (calculated on an $L = 48$ lattice with $N = 3 \times 48^2$ sites), the average magnetization along the $z$ axis $ M_z=\frac{1}{N}\sum_i \langle S_{i,z} \rangle$, the average total magnetic moment $M_{\mathrm{total}}=\frac{1}{N}\sum_i | \langle \boldsymbol{S_{i}} \rangle |$, and the DSC $\delta\chi$.
The specific phase boundaries are defined as follows:
\begin{itemize}
    \item \text{Skyrmion crystal (SkX) phase:} Characterized by a finite topological charge $\chi_Q \ge 3$, accompanied by a finite total magnetic moment $M_{\mathrm{total}} > 0.2$ and $z$-axis magnetization $M_z > 0.1$.
    \item \text{Fluctuating chiral (FC) phase:} Emerging primarily near the magnetic phase boundaries due to thermal excitations, this phase is characterized by an enhanced DSC, quantitatively defined by $\delta\chi > 0.2$.
    \item \text{Helical phase:} Identified by a vanishingly small topological charge $\chi_Q < 1.5$ and a small $z$-axis magnetization $M_z < 0.1$, but sustaining a total magnetic moment $M_{\mathrm{total}} > 0.15$.
    \item \text{Field-polarized (FP) phase:} Distinguished by strongly aligned spins, featuring a large $z$-axis magnetization $M_z > 0.9$ and total magnetic moment $M_{\mathrm{total}} > 0.7$, while the topological charge is largely suppressed ($\chi_Q < 3$).
    \item \text{Paramagnetic (Pm) phase:} Driven by thermal disorder, defined by the suppression of both the $z$-axis magnetization ($M_z < 0.2$) and the total magnetic moment ($M_{\mathrm{total}} < 0.1$).
\end{itemize}
Spin configurations that do not strictly satisfy any of the aforementioned criteria are classified as transition regions (TR).

\begin{figure}[tb]
  \centering
    \includegraphics[width=1\columnwidth]{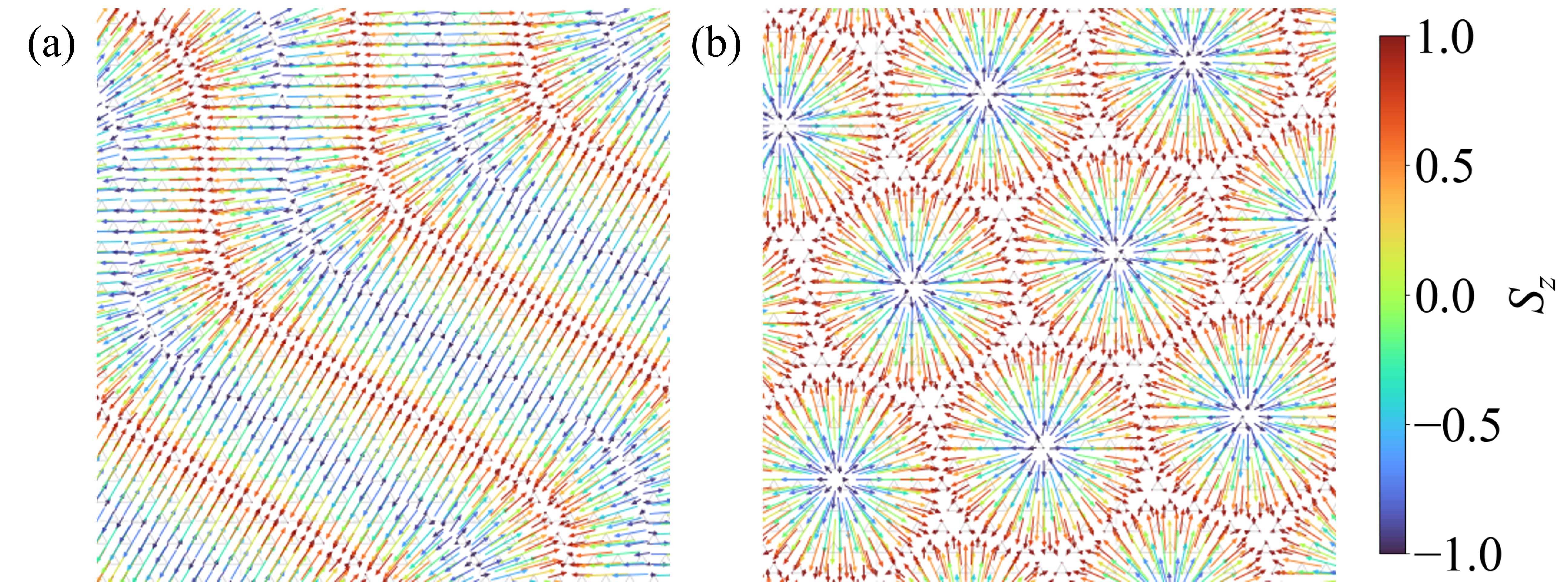}
    \caption{Snapshots of the real space spin configurations obtained from Monte Carlo simulations at a low temperature $T = 0.001 J_{\mathrm{H}}$.
    (a) A representative spin pattern at $B_z = 0.01 J_{\mathrm{H}}$, extracted from the full simulation lattice, showing the characteristic helical phase.
    (b) A representative spin pattern at $B_z = 0.068 J_{\mathrm{H}}$, showing a well defined SkX phase. In both panels, the arrows represent the local spin directions. The color scale indicates the out of plane spin component $S_z$.}
  \label{spin_pattern}
\end{figure}
Considering the presence of $\chi_Q$ in the SkX region and the DSC in the FC phase, as shown in Figs.~\ref{phase_diagram}(b) and \ref{phase_diagram}(c), we predict that these two areas will be dominated by intrinsic and extrinsic transport mechanisms, respectively.
This prediction is confirmed by the transport results presented later in the text.
To provide a microscopic basis for these phase boundaries, we display the representative real space spin patterns of the helical and SkX phases in Figs.~\ref{spin_pattern}(a) and \ref{spin_pattern}(b).
In the low field limit, local magnetic moments exhibit a periodic helical pattern.
At intermediate magnetic fields, the system transforms into a stable skyrmion lattice.
These real space textures corroborate the phase boundaries determined via topological charge and magnetization.

\begin{figure}[tb]
  \centering
    \includegraphics[width=1\columnwidth]{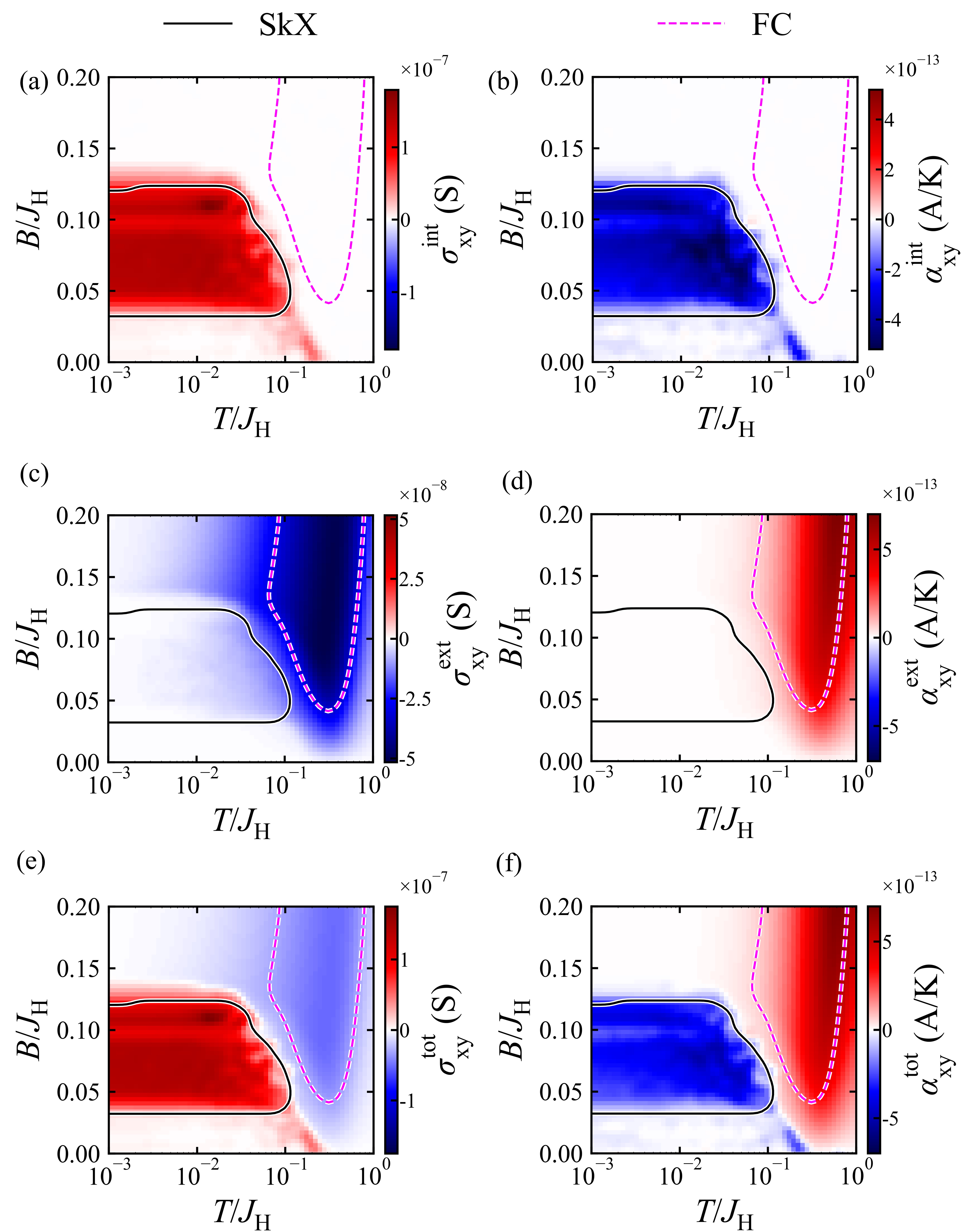}
    \caption{
    Distribution of the anomalous transport coefficients across the $T\text{-}B$ parameter space, calculated with $t = -1$~eV and $J_{\mathrm{K}} = 0.2t$.
    (a) Intrinsic AHC $\sigma_{xy}^{\mathrm{int}}$ and (b) intrinsic ANC $\alpha_{xy}^{\mathrm{int}}$, calculated by incorporating the thermally averaged static spin configurations $\langle \boldsymbol{S}_i \rangle$ from the MC simulations into an expanded $12 \times 12$ tight binding supercell.
    (c) Extrinsic AHC $\sigma_{xy}^{\mathrm{ext}}$ and (d) extrinsic ANC $\alpha_{xy}^{\mathrm{ext}}$ originating from the skew scattering mechanism, calculated using the analytical expressions driven by the dynamical chiral fluctuations $\delta \chi$.
    (e) Total AHC $\sigma_{xy}^{\mathrm{tot}}$ and (f) total ANC $\alpha_{xy}^{\mathrm{tot}}$, representing the summation of the respective intrinsic and extrinsic contributions.
    In all panels, the solid black curve and the dotted magenta curve denote the boundaries of the SkX phase and the fluctuating chiral phase, respectively, explicitly highlighting the spatial separation and competition of the intrinsic and extrinsic transport mechanisms.
    }
  \label{tans_phase}
\end{figure}

Utilizing the Monte Carlo results, intrinsic transport signatures driven by the static spin configurations are calculated.
To accommodate the spatial periodicity of magnetic textures, the electronic Hamiltonian is constructed on an expanded magnetic supercell.
Intrinsic AHC ($\sigma^{\mathrm{int}}_{xy}$) and ANC ($\alpha^{\mathrm{int}}_{xy}$) are calculated using Eqs.~\eqref{AHC_int} and \eqref{ANE_int}. 
To express the model results in physical units, we use representative energy scales $J_{\mathrm{H}} = 10$~meV and $t = -1$ eV, chosen to match the typical orders of magnitude of exchange and hopping energies in Kagome magnetic systems~\cite{Xie_Chen_2021,PhysRevB.105.035107}. 
The effective exchange coupling is set to $J_{\mathrm K}=0.2 t$, following the parameter choice in Ref.~\cite{Jsd0.2t}.
The Fermi energy is chosen as $\varepsilon_{\mathrm F}=-4.0$ eV to evaluate the transport response in the low energy region.
Distributions of these transport coefficients across the $T\text{-}B$ parameter space are mapped in Figs.~\ref{tans_phase}(a) and \ref{tans_phase}(b).
Intrinsic AHC and ANC emerge within the SkX phase in Figs.~\ref{phase_diagram}(d).
This intrinsic response is driven by the nonzero static scalar spin chirality.


Subsequently, extrinsic AHC and ANC induced by skew scattering are calculated by substituting spatially averaged DSC $\delta \chi$ into Eqs.~\eqref{AHC_ext_2d} and \eqref{ANE_ext_2d}.
Figs.~\ref{tans_phase}(c) and \ref{tans_phase}(d) illustrate these transport coefficients.
Within the FC phase between magnetically ordered and disordered phases, fluctuating scalar spin chirality in Fig.~\ref{phase_diagram}(c) generates extrinsic AHC ($\sigma^{\mathrm{ext}}_{xy}$) and ANC ($\alpha^{\mathrm{ext}}_{xy}$) via skew scattering.


Combining these intrinsic and extrinsic contributions yields the total AHC ($\sigma_{xy}^{\mathrm{tot}}$) and ANC ($\alpha_{xy}^{\mathrm{tot}}$), shown in Figs.~\ref{tans_phase}(e) and \ref{tans_phase}(f).
In the SkX regime, momentum space Berry curvature driven by static scalar spin chirality dominates the intrinsic response.
At the FC phase, thermal fluctuations suppress the intrinsic mechanism.
Within this phase, DSC dominates the AHC and ANC via skew scattering.
Calculations show that intrinsic and extrinsic mechanisms contribute with opposite signs.
This competition provides a microscopic explanation for the sign reversals of the topological Hall effect \cite{Nagaosa_2013} and topological Nernst effect \cite{Kolincio_2021} observed experimentally near phase transitions.

\section{Conclusions}
In this work, we investigate the interplay between static spin configurations and DSC in two dimensional Kagome lattices.
The combination of semiclassical Boltzmann transport theory and Monte Carlo simulations enables the decomposition of the total scalar spin chirality into static and dynamic components, from which the anomalous transport coefficients are obtained.
In the magnetically ordered SkX phase, static scalar spin chirality dominates the intrinsic anomalous Hall and Nernst responses.
Within the FC phase, DSC governs the transport via extrinsic skew scattering.
These mechanisms contribute with opposite signs, establishing a competition that determines the total anomalous Hall and Nernst response.
This mechanism clarifies anomalous transport in noncoplanar spin systems and informs the design of spintronic devices leveraging phase transitions.

\section {Acknowledgements}
This work is supported by the National Key R\&D Program of China (Grant No. 2024YFA1409200, No. 2022YFA1402802), CAS Project for Young Scientists in Basic Research Grant No. YSBR-057.  G.S. was supported in part by the Quantum Science and Technology-National Science and Technology Major Project under Grant No. 2024ZD0300500, NSFC Nos. 12534009 and 12447101, the Strategic Priority Research Program of CAS (Grant No. XDB1270000) and the CAS Superconducting Research Project under Grant No. SCZX-0101.

\appendix

\section{Derivation of the Skew Scattering Rate based on the Kondo Model} \label{app-A}

In this section, we provide a detailed derivation of the skew scattering rate within the framework of the Kondo model. The Kondo model characterizes the exchange coupling between itinerant conduction electrons and localized magnetic moments, providing a microscopic basis for spin-dependent scattering processes, such as asymmetric skew scattering. By employing perturbation theory and Fermi's golden rule, we formulate the scattering rate starting from the Kondo lattice Hamiltonian:
\begin{align}
    H &= H_0 + H_{\mathrm{K}} \nonumber \\
    &= \sum_{\langle ij \rangle, \sigma} t_{ij} c_{i\sigma}^{\dagger} c_{j\sigma} + J_{\mathrm{K}} \sum_{i} c_{i\alpha}^{\dagger} \boldsymbol{\sigma}_{\alpha\beta} c_{i\beta} \cdot \boldsymbol{S}_{i}, \label{Kondo_Hamilton_App}
\end{align}
where $H_0$ represents the kinetic energy of electrons in a tight binding representation and $H_{\mathrm{K}}$ denotes the Kondo exchange interaction.

To treat the scattering processes in momentum space, we define the Fourier transform of the electron operators:
\begin{align}
    c_{\sigma}(\boldsymbol{R}_i) &= \frac{1}{\sqrt{N}} \sum_{\boldsymbol{k}} e^{i \boldsymbol{k} \cdot \boldsymbol{R}_i} c_{\sigma}(\boldsymbol{k}), \\
    c_{\sigma}(\boldsymbol{k}) &= \frac{1}{\sqrt{N}} \sum_{i} e^{-i \boldsymbol{k} \cdot \boldsymbol{R}_i} c_{\sigma}(\boldsymbol{R}_i),
\end{align}
where $N$ is the number of unit cells. Substituting these into the Kondo term $H_{\mathrm{K}}$, the interaction Hamiltonian in the momentum representation is given by:
\begin{align}
    H_{\mathrm{K}} &= \frac{J_{\mathrm{K}}}{N} \sum_{i} \sum_{\boldsymbol{k}, \boldsymbol{k}^{\prime}} e^{-i (\boldsymbol{k}^{\prime} - \boldsymbol{k}) \cdot \boldsymbol{R}_i} \nonumber \\
    & \times \boldsymbol{S}_i \cdot \left( c_{\beta}^{\dagger}(\boldsymbol{k}^{\prime}) \boldsymbol{\sigma}_{\beta\alpha} c_{\alpha}(\boldsymbol{k}) \right). \label{H_K_momentum}
\end{align}

The transition matrix elements between the initial state $|\boldsymbol{k}\alpha\rangle$ and final state $|\boldsymbol{k}^{\prime}\beta\rangle$ are calculated using the T-matrix expansion. The first-order element is:
\begin{align}
    F_{\beta\alpha}^{(1)}(\boldsymbol{k}^{\prime}, \boldsymbol{k}) &= \langle \boldsymbol{k}^{\prime}\beta | H_{\mathrm{K}} | \boldsymbol{k}\alpha \rangle \nonumber \\
    &= \frac{J_{\mathrm{K}}}{N} \sum_{i} e^{-i (\boldsymbol{k}^{\prime}-\boldsymbol{k}) \cdot \boldsymbol{R}_i} \boldsymbol{S}_i \cdot \boldsymbol{\sigma}_{\beta\alpha}, \label{F_first_order}
\end{align}
and the second-order element is:
\begin{align}
    &F_{\beta\alpha}^{(2)}(\boldsymbol{k}^{\prime}, \boldsymbol{k}) = \langle \boldsymbol{k}^{\prime}\beta | H_{\mathrm{K}} G(\varepsilon_{\boldsymbol{k}}) H_{\mathrm{K}} | \boldsymbol{k}\alpha \rangle  \nonumber \\
    &= \left(\frac{J_{\mathrm{K}}}{N}\right)^2 \sum_{\boldsymbol{q}, \gamma} \frac{1}{\varepsilon_{\boldsymbol{k}} - \varepsilon_{\boldsymbol{q}} + i\eta} \label{F_second_order} \\
    & \times \sum_{i,j} (\boldsymbol{S}_i \cdot \boldsymbol{\sigma}_{\beta\gamma}) (\boldsymbol{S}_j \cdot \boldsymbol{\sigma}_{\gamma\alpha}) e^{i\boldsymbol{q}\cdot(\boldsymbol{R}_i-\boldsymbol{R}_j)} e^{i(\boldsymbol{k}\cdot \boldsymbol{R}_j - \boldsymbol{k}^{\prime}\cdot \boldsymbol{R}_i)}, \nonumber
\end{align}
where $G(\varepsilon_{\boldsymbol{k}}) = (\varepsilon_{\boldsymbol{k}} - H_0 + i\eta)^{-1}$ is the unperturbed retarded Green's function.

Under the low-energy approximation, we assume a parabolic dispersion $\varepsilon_{\boldsymbol{k}} = \hbar^2 k^2 / 2m$.
The summation over the intermediate momentum states $\boldsymbol{q}$ can then be performed analytically.
Using the spin algebraic identity $(\boldsymbol{S}_i \cdot \boldsymbol{\sigma}) (\boldsymbol{S}_j \cdot \boldsymbol{\sigma}) = \boldsymbol{S}_i \cdot \boldsymbol{S}_j \delta_{\beta\alpha} + i(\boldsymbol{S}_i \times \boldsymbol{S}_j) \cdot \boldsymbol{\sigma}_{\beta\alpha}$, we separate the scattering into spin independent and spin flip parts.
The second order matrix element is then expressed as:
\begin{align}
    F_{\beta\alpha}^{(2)}(\boldsymbol{k}^{\prime}, \boldsymbol{k}) & = -\left( \frac{m J_{\mathrm{K}}^2}{\hbar^2 N^2} \right) \frac{V}{2\pi} \nonumber \\
    & \times \sum_{i \neq j} \left[ \boldsymbol{S}_i \cdot \boldsymbol{S}_j \delta_{\beta\alpha} + i (\boldsymbol{S}_i \times \boldsymbol{S}_j) \cdot \boldsymbol{\sigma}_{\beta\alpha} \right] \nonumber \\
    & \times e^{i(\boldsymbol{k}\cdot\boldsymbol{R}_j - \boldsymbol{k}^{\prime}\cdot\boldsymbol{R}_i)} \left( \frac{e^{i k r_{ij}}}{r_{ij}} \right), \label{F_second_order_final}
\end{align}
where $r_{ij} = |\boldsymbol{R}_i - \boldsymbol{R}_j|$ is the distance between two localized moments, and $k = |\boldsymbol{k}|$ is the magnitude of the wave vector.

By substituting the first and second order matrix elements into Fermi's golden rule, the transition rate $W_{\boldsymbol{k}\alpha \to \boldsymbol{k}^{\prime}\beta}$ can be decomposed into a symmetric part $W^S$ and an antisymmetric part $w^-$.
The symmetric part $W^S$, primarily arising from the leading order Born approximation $|F^{(1)}|^2$, describes standard momentum relaxation.
In contrast, the antisymmetric part $w^-$, which satisfies $w_{\boldsymbol{k}\alpha \to \boldsymbol{k}^{\prime}\beta}^- = -w_{\boldsymbol{k}^{\prime}\beta \to \boldsymbol{k}\alpha}^-$, is responsible for the skew scattering and the resulting Hall transport.
This antisymmetric contribution originates from the interference between the first order exchange interaction and the second order transition process.
By isolating the terms that are odd under the exchange of $\boldsymbol{k}$ and $\boldsymbol{k}^{\prime}$ and retaining terms up to the third order in the Kondo coupling $J_{\mathrm{K}}$, we define the antisymmetric scattering rate as:
\begin{equation}
  2w_{\boldsymbol{k}\alpha \to \boldsymbol{k}^{\prime}\beta}^- = W_{\boldsymbol{k}\alpha \to \boldsymbol{k}^{\prime}\beta} - W_{\boldsymbol{k}^{\prime}\beta \to \boldsymbol{k}\alpha}. \label{w_minus_def}
\end{equation}
Following a detailed algebraic reduction of the phase factors and spin identities, the explicit form of the antisymmetric scattering rate is derived as:
\begin{align}
  &w_{\boldsymbol{k}\alpha \to \boldsymbol{k}^{\prime}\beta}^- = \frac{2\pi}{\hbar} \delta(\varepsilon_{\boldsymbol{k}} - \varepsilon_{\boldsymbol{k}^{\prime}}) \Bigg[ \frac{m J_{\mathrm{K}}^3 V}{2\pi \hbar^2 N^3} \sum_{l} \sum_{i \neq j} \nonumber \\
  &\quad \times (\boldsymbol{S}_l \cdot \boldsymbol{\sigma}_{\beta\alpha}) \left[ (\boldsymbol{S}_i \times \boldsymbol{S}_j) \cdot \boldsymbol{\sigma}_{\alpha\beta} \right] i \frac{e^{-i k r_{ij}}}{2r_{ij}} \nonumber \\
  &\quad \times \big( \cos[\boldsymbol{k} \cdot (\boldsymbol{R}_l - \boldsymbol{R}_j) - \boldsymbol{k}^{\prime} \cdot (\boldsymbol{R}_l - \boldsymbol{R}_i)] \nonumber \\
  &\quad - \cos[\boldsymbol{k} \cdot (\boldsymbol{R}_l - \boldsymbol{R}_i) - \boldsymbol{k}^{\prime} \cdot (\boldsymbol{R}_l - \boldsymbol{R}_j)] \big) + \mathrm{H.c.} \Bigg],
\end{align}
where $V$ is the total volume of the system, and $\mathrm{H.c.}$ denotes the Hermitian conjugate.

\section{Conductivity from Asymmetric Scattering}
To evaluate the macroscopic transport properties, we start from the Boltzmann equation under the relaxation time approximation for the symmetric part of the scattering.
The equation in the presence of an external electric field $\boldsymbol{E}$ is given by:
\begin{align}
  &-e (\boldsymbol{v}_{\boldsymbol{k},\alpha} \cdot \boldsymbol{E}) f_0^{\prime}(\varepsilon_{\boldsymbol{k},\alpha}) \nonumber \\
  &\quad = -\frac{g_{\boldsymbol{k}}^\alpha}{\tau} + \sum_{\beta} \int \frac{V d^3 \boldsymbol{k}^{\prime}}{(2\pi)^3} w_{\boldsymbol{k}^{\prime}\beta \to \boldsymbol{k}\alpha}^{-} g_{\boldsymbol{k}^{\prime}}^\beta, \label{Boltzmann_eq_app}
\end{align}
where $g_{\boldsymbol{k}}^\alpha$ represents the nonequilibrium part of the distribution function, and $\tau$ is the transport relaxation time.
The second term on the right-hand side represents the collision integral specifically arising from the antisymmetric scattering rate $w^{-}$.
By transforming the integral into spherical coordinates, Eq.~\eqref{Boltzmann_eq_app} becomes:
\begin{align}
  &-e (\boldsymbol{v}_{\boldsymbol{k},\alpha} \cdot \boldsymbol{E}) f_0^{\prime}(\varepsilon_{\boldsymbol{k},\alpha}) = -\frac{g_{\boldsymbol{k}}^\alpha}{\tau} \nonumber \\
  &\quad + \sum_{\beta} \frac{V}{(2\pi)^3} \int (k^{\prime})^2 \sin\theta^{\prime} d\theta^{\prime} d\phi^{\prime} dk^{\prime} \, w_{\boldsymbol{k}^{\prime}\beta \to \boldsymbol{k}\alpha}^{-} g_{\boldsymbol{k}^{\prime}}^\beta.
\end{align}
We assume that the antisymmetric scattering rate takes the following form, which characterizes the skew scattering in magnetic systems:
\begin{equation}
  w_{\boldsymbol{k}^{\prime}\beta \to \boldsymbol{k}\alpha}^- = \tilde{\boldsymbol{V}}_{\alpha\beta}(k) \cdot \frac{\boldsymbol{k} \times \boldsymbol{k}^{\prime}}{k^2} \delta(\varepsilon_{\boldsymbol{k}\alpha} - \varepsilon_{\boldsymbol{k}^{\prime}\beta}),
\end{equation}
where $\tilde{\boldsymbol{V}}_{\alpha\beta}(k)$ is a spin dependent vector characterizing the strength of the asymmetric interaction.
To simplify the collision integral, we introduce the auxiliary vector function:
\begin{equation}
  \boldsymbol{P}_{\beta}(k^{\prime}) = \int \sin\theta^{\prime} d\theta^{\prime} d\phi^{\prime} \boldsymbol{k}^{\prime} g_{\boldsymbol{k}^{\prime}}^\beta |_{\varepsilon_{\boldsymbol{k}\alpha} = \varepsilon_{\boldsymbol{k}^{\prime}\beta}}.
\end{equation}
Utilizing this auxiliary function and performing the integration over the energy Dirac delta function, the Boltzmann equation simplifies to:
\begin{align}
  g_{\boldsymbol{k}}^\alpha &= \tau e (\boldsymbol{v}_{\boldsymbol{k},\alpha} \cdot \boldsymbol{E}) f_0^{\prime}(\varepsilon_{\boldsymbol{k},\alpha}) \nonumber \\
  & + \sum_{\beta} V \tau \frac{\rho(k)}{4\pi} \tilde{\boldsymbol{V}}_{\alpha\beta}(k) \cdot \frac{\boldsymbol{k} \times \boldsymbol{P}_{\beta}(k^{\prime})}{k^2},
\end{align}
where $\rho(\varepsilon_{\boldsymbol{k}}) = \frac{k^2}{2\pi^2 |d\varepsilon/dk|} = \frac{mk}{2\pi^2 \hbar^2}$ is the density of states per spin per unit volume.

To account for the spin structure of the scattering, we consider the identity for the product of spin matrices:
\begin{equation}
  (\boldsymbol{\nu} \cdot \boldsymbol{\sigma}_{\alpha\beta}) (\boldsymbol{\sigma}_{\beta\alpha} \cdot \boldsymbol{\nu}^{\prime}) =
  \begin{cases}
    \nu_z \nu_z^{\prime}, & \text{if } \alpha = \beta \\
    \nu_x \nu_x^{\prime} + \nu_y \nu_y^{\prime}, & \text{if } \alpha \neq \beta
  \end{cases}
\end{equation}
Based on this structure, we decompose the interaction vector $\tilde{\boldsymbol{V}}_{\alpha\beta}(k)$ along the spatial direction $\boldsymbol{n}$ using the Pauli matrices $\sigma_0$ and $\sigma_{x,y}$ in the spin space:
\begin{equation}
  \tilde{\boldsymbol{V}}_{\alpha\beta}(k) = \left( \tilde{V}^0(k) \sigma_0^{\alpha\beta} + \tilde{V}^1(k) \sigma_x^{\alpha\beta} + i \tilde{V}^2(k) \sigma_y^{\alpha\beta} \right) \boldsymbol{n}. \label{auxi_V_app}
\end{equation}
Explicitly, the components for different spin channels are:
\begin{align}
\tilde{\boldsymbol{V}}_{\uparrow\uparrow}(k) &= \tilde{V}^0(k) \boldsymbol{n}, \\
\tilde{\boldsymbol{V}}_{\downarrow\downarrow}(k) &= \tilde{V}^0(k) \boldsymbol{n}, \\
\tilde{\boldsymbol{V}}_{\uparrow\downarrow}(k) &= (\tilde{V}^1(k) + \tilde{V}^2(k)) \boldsymbol{n}, \\
\tilde{\boldsymbol{V}}_{\downarrow\uparrow}(k) &= (\tilde{V}^1(k) - \tilde{V}^2(k)) \boldsymbol{n}.
\end{align}
Finally, to parameterize the spin flip scattering asymmetry, we define the parameter $\eta$ through $e^{-\eta} = (\tilde{V}^1(k) + \tilde{V}^2(k)) / \sqrt{(\tilde{V}^1(k))^2 - (\tilde{V}^2(k))^2}$.

By solving the linearized Boltzmann equation for the spin dependent distribution functions $g^{\uparrow}_{\boldsymbol{k}}$ and $g^{\downarrow}_{\boldsymbol{k}}$ to first order in the electric field $\boldsymbol{E}$, we can extract the transverse Hall current density $j_x$.
For an electric field applied along the $y$ direction ($\boldsymbol{E} = E_y \hat{\boldsymbol{y}}$) and the interaction vector oriented with a $z$ component $n_z$, the Hall current density is given by:
\begin{align}
  j_x &= -e^2 V\tau^2 \int E_{y} \frac{\rho(\varepsilon_{k})}{9} \frac{k^2 dk}{2\pi^2} \nonumber \\
      &\quad \times \left[ 2\left(\tilde{V}^0_z(k) + \tilde{V}^1_z(k)\right) \frac{\partial f_0}{\partial \varepsilon} \right] v_{k_{x}}^2, \nonumber\\
      &= e^2 V \tau^2 \frac{k_{\mathrm{F}} m}{9\pi^2 \hbar^2} v_{k_{x,\mathrm{F}}}^2 \rho(\varepsilon_{k_{\mathrm{F}}}) \nonumber \\
      &\quad \times \left[ \left(\tilde{V}^0(k_{\mathrm{F}}) + \tilde{V}^1(k_{\mathrm{F}})\right) E_y n_z \right]. \label{j_x_AHC_app}
\end{align}
From the definition of the Hall conductivity $\sigma_{xy} = j_x / E_y$, and focusing on the $z$ components $\tilde{V}^0_z$ and $\tilde{V}^1_z$, we obtain:
\begin{align}
  \sigma_{xy} &= e^2 V \tau^2 \frac{k_{\mathrm{F}}^3}{9\pi^2 m} \rho(\varepsilon_{k_{\mathrm{F}}}) \left( \tilde{V}_z^0(k_{\mathrm{F}}) + \tilde{V}_z^1(k_{\mathrm{F}}) \right). \label{sigma_xy1_app}
\end{align}
Finally, by substituting the 3D carrier density per spin species $n_e = k_{\mathrm{F}}^3 / (6\pi^2)$, we obtain the simplified expression for the extrinsic anomalous Hall conductivity:
\begin{equation}
  \sigma_{xy}^{\mathrm{ext}} = \frac{2 n_e e^2 \tau^2}{3 m} V \rho(\varepsilon_{\mathrm{F}}) \left( \tilde{V}_z^0(k_{\mathrm{F}}) + \tilde{V}_z^1(k_{\mathrm{F}}) \right).
\end{equation}

\section{Anomalous Nernst Conductivity and Verification of the Mott Relation} \label{app-C}
To evaluate the anomalous Nernst conductivity (ANC), we start from the Boltzmann equation in the presence of an in plane temperature gradient $\nabla T$.
Treating the temperature gradient as a statistical driving force, the linearized Boltzmann equation under the relaxation time approximation is given by:
\begin{align}
    &\left( \boldsymbol{v}_{\boldsymbol{k},\alpha} \cdot \nabla T \right) \frac{\partial f_0(\varepsilon_{\boldsymbol{k},\alpha})}{\partial T} \nonumber \\
    &\quad = -\frac{g_{\boldsymbol{k}}^\alpha}{\tau} + \sum_{\beta} \frac{V}{(2\pi)^{3}} \int d^3 \boldsymbol{k}^{\prime} w_{\boldsymbol{k}^{\prime}\beta \to \boldsymbol{k}\alpha}^{-} g_{\boldsymbol{k}^{\prime}}^\beta, \nonumber \\
    &\quad = -\frac{g_{\boldsymbol{k}}^\alpha}{\tau} + \sum_{\beta} \frac{V}{(2\pi)^{3}} \int (k^{\prime})^2 \sin\theta^{\prime} \, dk^{\prime} \, d\theta^{\prime} \, d\phi^{\prime} \nonumber \\
    &\quad\quad \times w_{\boldsymbol{k}^{\prime}\beta \to \boldsymbol{k}\alpha}^{-} g_{\boldsymbol{k}^{\prime}}^\beta. \label{Boltzmann_ANC_app}
\end{align}
Because the collision integral on the right-hand side is mathematically identical to that of the electrical transport case, solving this integral equation directly parallels the derivation of the AHC. The essential difference is that the electrical driving term $e\boldsymbol{v} \cdot \boldsymbol{E} (\partial f_0 / \partial \varepsilon)$ is replaced by the thermal driving term $-\boldsymbol{v} \cdot \nabla T (\partial f_0 / \partial T)$.
By drawing this direct analogy, we can explicitly construct the transverse charge current $j_x$ driven by a longitudinal temperature gradient ($\nabla T = \nabla_y T \hat{\boldsymbol{y}}$). Integrating over the momentum space and substituting the asymmetric scattering potentials, we obtain:
\begin{align}
    j_x &= e V \tau^2 \int \frac{k^2 dk}{(2\pi)^3} \frac{\rho(k)}{6} \left( \frac{4\pi}{3} v_{{k}_{x}}^2 \right) \nonumber \\
    &\quad \times \left[ \nabla_y T \, {n}_z \left( 4\tilde{V}^0(k) + 4\tilde{V}^1(k) \right) \frac{\partial f_0}{\partial T} \right].
\end{align}

To verify the validity of the Mott relation, we define an energy dependent anomalous Hall coefficient:
\begin{equation}
  \varsigma_{xy}(\varepsilon_k) = -e^2 V \tau^2 \frac{k^3}{9\pi^2 m} \rho(\varepsilon_{k}) \left( \tilde{V}_z^0(k) + \tilde{V}_z^1(k) \right).
  \label{ANC_explicit_integral}
\end{equation}
By inserting this definition, the Nernst conductivity $\alpha_{xy} = j_x / (-\nabla_y T)$ can be rewritten as an energy integral:
\begin{equation}
    \alpha_{xy} = \frac{1}{e} \int d\varepsilon \, \varsigma_{xy}(\varepsilon) \frac{\partial f_0}{\partial T}.
\end{equation}
Using the thermodynamic identity for the Fermi-Dirac distribution, $\frac{\partial f_0}{\partial T} = - \left( \frac{\varepsilon - \varepsilon_{\mathrm{F}}}{T} \right) \frac{\partial f_0}{\partial \varepsilon}$, the integral becomes:
\begin{equation}
    \alpha_{xy} = \frac{1}{e T} \int d\varepsilon \, \varsigma_{xy}(\varepsilon) (\varepsilon - \varepsilon_{\mathrm{F}}) \left( -\frac{\partial f_0}{\partial \varepsilon} \right).
\end{equation}
At low temperatures ($k_{\mathrm{B}} T \ll \varepsilon_{\mathrm{F}}$), the derivative $(-\partial f_0 / \partial \varepsilon)$ is strongly peaked at the Fermi energy $\varepsilon_{\mathrm{F}}$. Applying the Sommerfeld expansion, $\int d\varepsilon H(\varepsilon) (-\partial f_0 / \partial \varepsilon) \approx H(\varepsilon_{\mathrm{F}}) + \frac{\pi^2}{6}(k_{\mathrm{B}} T)^2 H^{\prime\prime}(\varepsilon_{\mathrm{F}})$, and defining the auxiliary function $H(\varepsilon) = \varsigma_{xy}(\varepsilon)(\varepsilon - \varepsilon_{\mathrm{F}})$, we find that $H(\varepsilon_{\mathrm{F}}) = 0$ and $H^{\prime\prime}(\varepsilon_{\mathrm{F}}) = 2 \left. \frac{\partial \varsigma_{xy}(\varepsilon)}{\partial \varepsilon} \right|_{\varepsilon_{\mathrm{F}}}$.

Consequently, the thermal integration evaluates exactly to:
\begin{align}
    \alpha_{xy}^{\mathrm{ext}} &= \frac{1}{e T} \frac{\pi^2}{6} (k_{\mathrm{B}} T)^2 \left( 2 \left. \frac{\partial \varsigma_{xy}(\varepsilon)}{\partial \varepsilon} \right|_{\varepsilon = \varepsilon_{\mathrm{F}}} \right) \nonumber \\
    &= -\frac{\pi^2 k_{\mathrm{B}}^2 T}{3e} \left. \frac{\partial \sigma_{xy}^{\mathrm{ext}}(\varepsilon)}{\partial \varepsilon} \right|_{\varepsilon = \varepsilon_{\mathrm{F}}}.
\end{align}
This rigorous transformation proves that our explicit evaluation of the transverse thermal response perfectly recovers the semiclassical Mott relation.

\onecolumngrid
\vspace{1em} 

\section{Extended Monte Carlo Simulation Results}

\begin{figure}[h!]
  \centering
  \includegraphics[width=\textwidth]{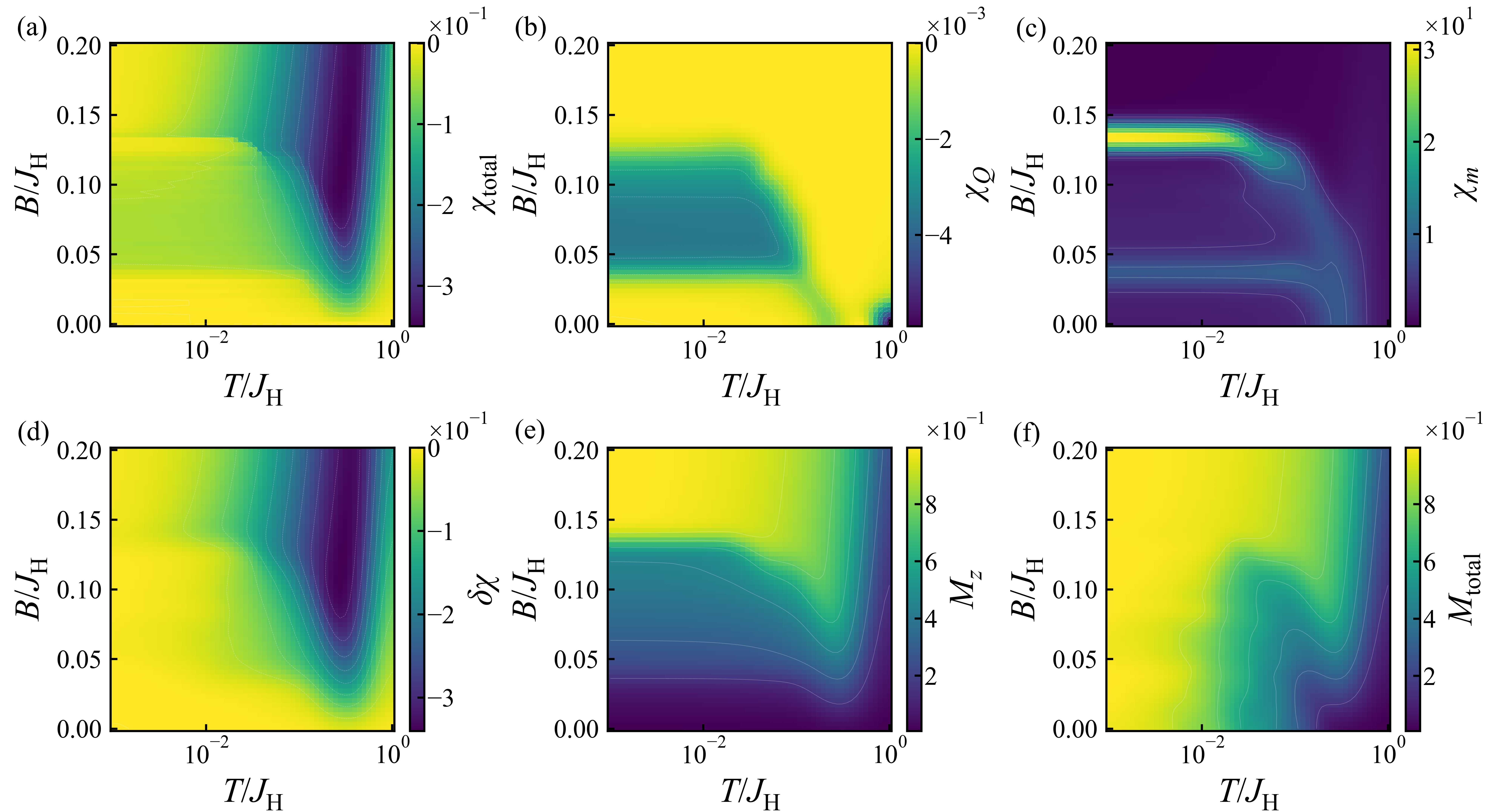}
  \caption{Phase diagrams of thermodynamic and topological quantities calculated via Monte Carlo simulations on a $60 \times 60$ lattice.
  (a) Total scalar spin chirality $\chi_{\mathrm{total}}$.
  (b) Topological charge $\chi_Q$.
  (c) Magnetic susceptibility $\chi_m$.
  (d) Fluctuation induced scalar spin chirality $\delta\chi$.
  (e) Average magnetization along the $z$ axis $M_z$.
  (f) Distribution of the average total magnetic moment $M_{\mathrm{total}}$.}
  \label{fig_MC_results_app}
\end{figure}

\begin{figure}[tb]
  \centering
  \includegraphics[width=\textwidth]{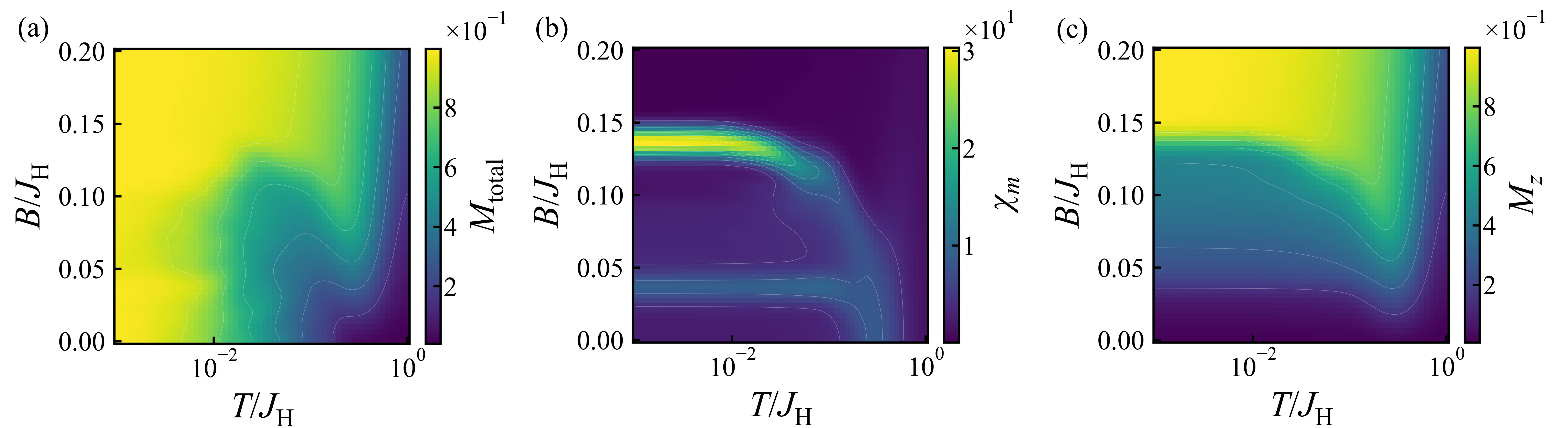}
  \caption{Thermodynamic phase diagrams as a function of temperature $T/J_{\mathrm{H}}$ and external magnetic field $B_z/J_{\mathrm{H}}$, obtained from Monte Carlo simulations on a $48 \times 48$ lattice.
  (a) Average magnitude of the average total magnetic moment $M_{\mathrm{total}}$.
  (b) Magnetic susceptibility $\chi_m$.
  (c) Average magnetization along the $z$-axis $M_z$.}
  \label{fig_phase_diagrams_app}
\end{figure}

\begin{figure}[tb]
  \centering
  \includegraphics[width=\textwidth]{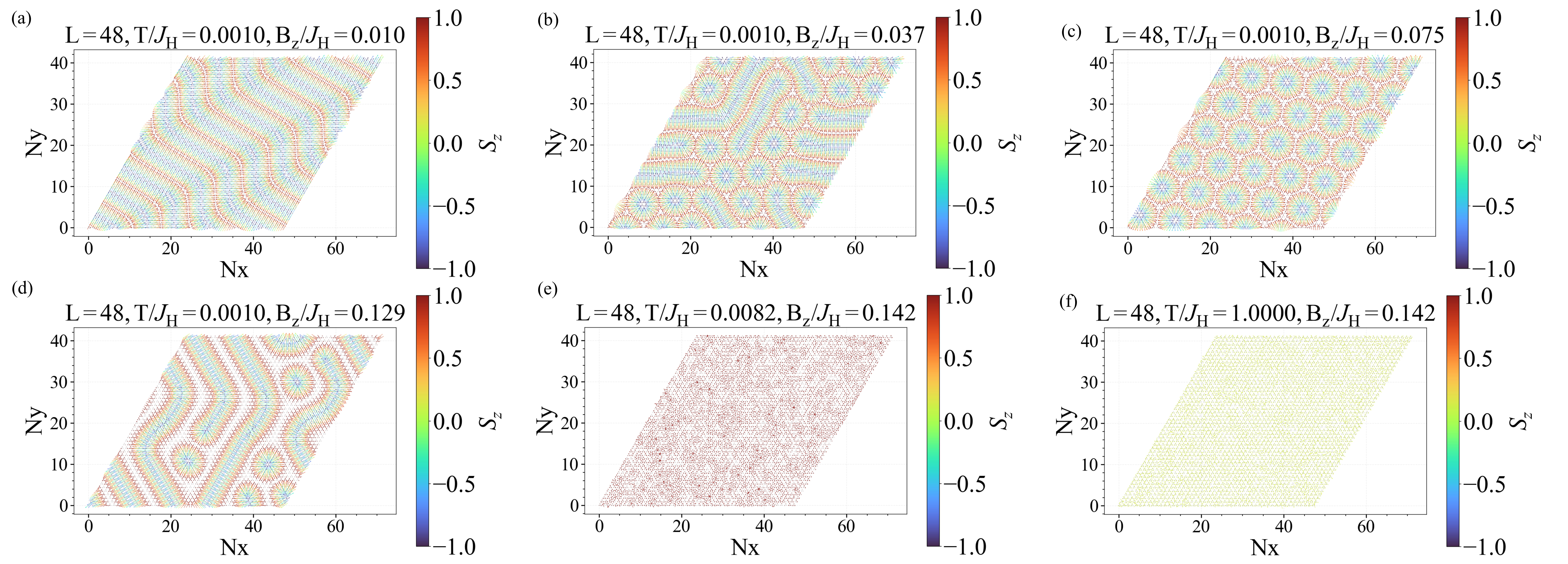}
  \caption{Real space spin configurations obtained from Monte Carlo simulations on a finite lattice ($L=48$) under various temperatures ($T/J_{\mathrm{H}}$) and out of plane magnetic fields ($B_z/J_{\mathrm{H}}$).
  The color map represents the out of plane spin component $S_z$, while the small arrows indicate the in plane spin orientations.
  (a) The helical phase at a low magnetic field ($T/J_{\mathrm{H}} = 0.001, B_z/J_{\mathrm{H}} = 0.010$).
  (b) The transition phase between the helical and skyrmion phases ($T/J_{\mathrm{H}} = 0.001, B_z/J_{\mathrm{H}} = 0.037$).
  (c) The SkX phase ($T/J_{\mathrm{H}} = 0.001, B_z/J_{\mathrm{H}} = 0.075$).
  (d) The transition phase between the skyrmion and field-polarized phases ($T/J_{\mathrm{H}} = 0.001, B_z/J_{\mathrm{H}} = 0.129$).
  (e) The field polarized phase at a high magnetic field ($T/J_{\mathrm{H}} = 0.008, B_z/J_{\mathrm{H}} = 0.142$).
  (f) The paramagnetic phase at a high temperature ($T/J_{\mathrm{H}} = 1.000, B_z/J_{\mathrm{H}} = 0.142$).}
  \label{fig_spin_textures_app}
\end{figure}

\clearpage
\twocolumngrid


\begin{thebibliography}{43}%
\makeatletter
\providecommand \@ifxundefined [1]{%
 \@ifx{#1\undefined}
}%
\providecommand \@ifnum [1]{%
 \ifnum #1\expandafter \@firstoftwo
 \else \expandafter \@secondoftwo
 \fi
}%
\providecommand \@ifx [1]{%
 \ifx #1\expandafter \@firstoftwo
 \else \expandafter \@secondoftwo
 \fi
}%
\providecommand \natexlab [1]{#1}%
\providecommand \enquote  [1]{``#1''}%
\providecommand \bibnamefont  [1]{#1}%
\providecommand \bibfnamefont [1]{#1}%
\providecommand \citenamefont [1]{#1}%
\providecommand \href@noop [0]{\@secondoftwo}%
\providecommand \href [0]{\begingroup \@sanitize@url \@href}%
\providecommand \@href[1]{\@@startlink{#1}\@@href}%
\providecommand \@@href[1]{\endgroup#1\@@endlink}%
\providecommand \@sanitize@url [0]{\catcode `\\12\catcode `\$12\catcode `\&12\catcode `\#12\catcode `\^12\catcode `\_12\catcode `\%12\relax}%
\providecommand \@@startlink[1]{}%
\providecommand \@@endlink[0]{}%
\providecommand \url  [0]{\begingroup\@sanitize@url \@url }%
\providecommand \@url [1]{\endgroup\@href {#1}{\urlprefix }}%
\providecommand \urlprefix  [0]{URL }%
\providecommand \Eprint [0]{\href }%
\providecommand \doibase [0]{https://doi.org/}%
\providecommand \selectlanguage [0]{\@gobble}%
\providecommand \bibinfo  [0]{\@secondoftwo}%
\providecommand \bibfield  [0]{\@secondoftwo}%
\providecommand \translation [1]{[#1]}%
\providecommand \BibitemOpen [0]{}%
\providecommand \bibitemStop [0]{}%
\providecommand \bibitemNoStop [0]{.\EOS\space}%
\providecommand \EOS [0]{\spacefactor3000\relax}%
\providecommand \BibitemShut  [1]{\csname bibitem#1\endcsname}%
\let\auto@bib@innerbib\@empty
\bibitem [{\citenamefont {Balents}(2010)}]{Balents_2010}%
  \BibitemOpen
  \bibfield  {author} {\bibinfo {author} {\bibfnamefont {L.}~\bibnamefont {Balents}},\ }\bibfield  {title} {\bibinfo {title} {Spin liquids in frustrated magnets},\ }\href {https://doi.org/10.1038/nature08917} {\bibfield  {journal} {\bibinfo  {journal} {Nature}\ }\textbf {\bibinfo {volume} {464}},\ \bibinfo {pages} {199} (\bibinfo {year} {2010})}\BibitemShut {NoStop}%
\bibitem [{\citenamefont {Yin}\ \emph {et~al.}(2022)\citenamefont {Yin}, \citenamefont {Lian},\ and\ \citenamefont {Hasan}}]{Yin_2022}%
  \BibitemOpen
  \bibfield  {author} {\bibinfo {author} {\bibfnamefont {J.-X.}\ \bibnamefont {Yin}}, \bibinfo {author} {\bibfnamefont {B.}~\bibnamefont {Lian}},\ and\ \bibinfo {author} {\bibfnamefont {M.~Z.}\ \bibnamefont {Hasan}},\ }\bibfield  {title} {\bibinfo {title} {Topological kagome magnets and superconductors},\ }\href {https://doi.org/10.1038/s41586-022-05516-0} {\bibfield  {journal} {\bibinfo  {journal} {Nature}\ }\textbf {\bibinfo {volume} {612}},\ \bibinfo {pages} {647} (\bibinfo {year} {2022})}\BibitemShut {NoStop}%
\bibitem [{\citenamefont {Nagaosa}\ \emph {et~al.}(2010)\citenamefont {Nagaosa}, \citenamefont {Sinova}, \citenamefont {Onoda}, \citenamefont {MacDonald},\ and\ \citenamefont {Ong}}]{Nagaosa_2010}%
  \BibitemOpen
  \bibfield  {author} {\bibinfo {author} {\bibfnamefont {N.}~\bibnamefont {Nagaosa}}, \bibinfo {author} {\bibfnamefont {J.}~\bibnamefont {Sinova}}, \bibinfo {author} {\bibfnamefont {S.}~\bibnamefont {Onoda}}, \bibinfo {author} {\bibfnamefont {A.~H.}\ \bibnamefont {MacDonald}},\ and\ \bibinfo {author} {\bibfnamefont {N.~P.}\ \bibnamefont {Ong}},\ }\bibfield  {title} {\bibinfo {title} {Anomalous {Hall} effect},\ }\href {https://doi.org/10.1103/RevModPhys.82.1539} {\bibfield  {journal} {\bibinfo  {journal} {Rev. Mod. Phys.}\ }\textbf {\bibinfo {volume} {82}},\ \bibinfo {pages} {1539} (\bibinfo {year} {2010})}\BibitemShut {NoStop}%
\bibitem [{\citenamefont {Sakai}\ \emph {et~al.}(2018)\citenamefont {Sakai}, \citenamefont {Mizuta}, \citenamefont {Nugroho}, \citenamefont {Sihombing}, \citenamefont {Koretsune}, \citenamefont {Suzuki}, \citenamefont {Takemori}, \citenamefont {Ishii}, \citenamefont {Nishio-Hamane}, \citenamefont {Arita}, \citenamefont {Goswami},\ and\ \citenamefont {Nakatsuji}}]{Sakai_2018}%
  \BibitemOpen
  \bibfield  {author} {\bibinfo {author} {\bibfnamefont {A.}~\bibnamefont {Sakai}}, \bibinfo {author} {\bibfnamefont {Y.~P.}\ \bibnamefont {Mizuta}}, \bibinfo {author} {\bibfnamefont {A.~A.}\ \bibnamefont {Nugroho}}, \bibinfo {author} {\bibfnamefont {R.}~\bibnamefont {Sihombing}}, \bibinfo {author} {\bibfnamefont {T.}~\bibnamefont {Koretsune}}, \bibinfo {author} {\bibfnamefont {M.-T.}\ \bibnamefont {Suzuki}}, \bibinfo {author} {\bibfnamefont {N.}~\bibnamefont {Takemori}}, \bibinfo {author} {\bibfnamefont {R.}~\bibnamefont {Ishii}}, \bibinfo {author} {\bibfnamefont {D.}~\bibnamefont {Nishio-Hamane}}, \bibinfo {author} {\bibfnamefont {R.}~\bibnamefont {Arita}}, \bibinfo {author} {\bibfnamefont {P.}~\bibnamefont {Goswami}},\ and\ \bibinfo {author} {\bibfnamefont {S.}~\bibnamefont {Nakatsuji}},\ }\bibfield  {title} {\bibinfo {title} {Giant anomalous {Nernst} effect and quantum-critical scaling in a ferromagnetic semimetal},\ }\href {https://doi.org/10.1038/s41567-018-0225-6} {\bibfield  {journal} {\bibinfo  {journal} {Nat. Phys.}\ }\textbf {\bibinfo {volume} {14}},\ \bibinfo {pages} {1119} (\bibinfo {year} {2018})}\BibitemShut {NoStop}%
\bibitem [{\citenamefont {Chen}\ \emph {et~al.}(2022)\citenamefont {Chen}, \citenamefont {Minami}, \citenamefont {Sakai}, \citenamefont {Wang}, \citenamefont {Feng}, \citenamefont {Nomoto}, \citenamefont {Hirayama}, \citenamefont {Ishii}, \citenamefont {Koretsune}, \citenamefont {Arita},\ and\ \citenamefont {Nakatsuji}}]{Chen_2022}%
  \BibitemOpen
  \bibfield  {author} {\bibinfo {author} {\bibfnamefont {T.}~\bibnamefont {Chen}}, \bibinfo {author} {\bibfnamefont {S.}~\bibnamefont {Minami}}, \bibinfo {author} {\bibfnamefont {A.}~\bibnamefont {Sakai}}, \bibinfo {author} {\bibfnamefont {Y.}~\bibnamefont {Wang}}, \bibinfo {author} {\bibfnamefont {Z.}~\bibnamefont {Feng}}, \bibinfo {author} {\bibfnamefont {T.}~\bibnamefont {Nomoto}}, \bibinfo {author} {\bibfnamefont {M.}~\bibnamefont {Hirayama}}, \bibinfo {author} {\bibfnamefont {R.}~\bibnamefont {Ishii}}, \bibinfo {author} {\bibfnamefont {T.}~\bibnamefont {Koretsune}}, \bibinfo {author} {\bibfnamefont {R.}~\bibnamefont {Arita}},\ and\ \bibinfo {author} {\bibfnamefont {S.}~\bibnamefont {Nakatsuji}},\ }\bibfield  {title} {\bibinfo {title} {Large anomalous {Nernst} effect and nodal plane in an iron-based kagome ferromagnet},\ }\href {https://doi.org/10.1126/sciadv.abk1480} {\bibfield  {journal} {\bibinfo  {journal} {Sci. Adv.}\ }\textbf {\bibinfo {volume} {8}},\ \bibinfo {pages} {eabk1480} (\bibinfo {year} {2022})}\BibitemShut {NoStop}%
\bibitem [{\citenamefont {Mizuguchi}\ and\ \citenamefont {Nakatsuji}(2019)}]{Mizuguchi_2019}%
  \BibitemOpen
  \bibfield  {author} {\bibinfo {author} {\bibfnamefont {M.}~\bibnamefont {Mizuguchi}}\ and\ \bibinfo {author} {\bibfnamefont {S.}~\bibnamefont {Nakatsuji}},\ }\bibfield  {title} {\bibinfo {title} {Energy-harvesting materials based on the anomalous {Nernst} effect},\ }\href {https://doi.org/10.1080/14686996.2019.1585143} {\bibfield  {journal} {\bibinfo  {journal} {Sci. Technol. Adv. Mater.}\ }\textbf {\bibinfo {volume} {20}},\ \bibinfo {pages} {262} (\bibinfo {year} {2019})}\BibitemShut {NoStop}%
\bibitem [{\citenamefont {Xiao}\ \emph {et~al.}(2010)\citenamefont {Xiao}, \citenamefont {Chang},\ and\ \citenamefont {Niu}}]{Xiao_2010}%
  \BibitemOpen
  \bibfield  {author} {\bibinfo {author} {\bibfnamefont {D.}~\bibnamefont {Xiao}}, \bibinfo {author} {\bibfnamefont {M.-C.}\ \bibnamefont {Chang}},\ and\ \bibinfo {author} {\bibfnamefont {Q.}~\bibnamefont {Niu}},\ }\bibfield  {title} {\bibinfo {title} {{Berry} phase effects on electronic properties},\ }\href {https://doi.org/10.1103/RevModPhys.82.1959} {\bibfield  {journal} {\bibinfo  {journal} {Rev. Mod. Phys.}\ }\textbf {\bibinfo {volume} {82}},\ \bibinfo {pages} {1959} (\bibinfo {year} {2010})}\BibitemShut {NoStop}%
\bibitem [{\citenamefont {Tatara}\ and\ \citenamefont {Kawamura}(2002)}]{Tatara_2002}%
  \BibitemOpen
  \bibfield  {author} {\bibinfo {author} {\bibfnamefont {G.}~\bibnamefont {Tatara}}\ and\ \bibinfo {author} {\bibfnamefont {H.}~\bibnamefont {Kawamura}},\ }\bibfield  {title} {\bibinfo {title} {Chirality-driven anomalous {Hall} effect in weak coupling regime},\ }\href {https://doi.org/10.1143/JPSJ.71.2613} {\bibfield  {journal} {\bibinfo  {journal} {J. Phys. Soc. Jpn.}\ }\textbf {\bibinfo {volume} {71}},\ \bibinfo {pages} {2613} (\bibinfo {year} {2002})}\BibitemShut {NoStop}%
\bibitem [{\citenamefont {Onoda}\ and\ \citenamefont {Nagaosa}(2003)}]{Nagaosa_2003}%
  \BibitemOpen
  \bibfield  {author} {\bibinfo {author} {\bibfnamefont {S.}~\bibnamefont {Onoda}}\ and\ \bibinfo {author} {\bibfnamefont {N.}~\bibnamefont {Nagaosa}},\ }\bibfield  {title} {\bibinfo {title} {Spin chirality fluctuations and anomalous {Hall} effect in itinerant ferromagnets},\ }\href {https://doi.org/10.1103/PhysRevLett.90.196602} {\bibfield  {journal} {\bibinfo  {journal} {Phys. Rev. Lett.}\ }\textbf {\bibinfo {volume} {90}},\ \bibinfo {pages} {196602} (\bibinfo {year} {2003})}\BibitemShut {NoStop}%
\bibitem [{\citenamefont {Bruno}\ \emph {et~al.}(2004)\citenamefont {Bruno}, \citenamefont {Dugaev},\ and\ \citenamefont {Taillefumier}}]{Bruno_2004}%
  \BibitemOpen
  \bibfield  {author} {\bibinfo {author} {\bibfnamefont {P.}~\bibnamefont {Bruno}}, \bibinfo {author} {\bibfnamefont {V.~K.}\ \bibnamefont {Dugaev}},\ and\ \bibinfo {author} {\bibfnamefont {M.}~\bibnamefont {Taillefumier}},\ }\bibfield  {title} {\bibinfo {title} {Topological {Hall} effect and {Berry} phase in magnetic nanostructures},\ }\href {https://doi.org/10.1103/PhysRevLett.93.096806} {\bibfield  {journal} {\bibinfo  {journal} {Phys. Rev. Lett.}\ }\textbf {\bibinfo {volume} {93}},\ \bibinfo {pages} {096806} (\bibinfo {year} {2004})}\BibitemShut {NoStop}%
\bibitem [{\citenamefont {Nakazawa}\ \emph {et~al.}(2018)\citenamefont {Nakazawa}, \citenamefont {Bibes},\ and\ \citenamefont {Kohno}}]{Nakazawa_2018}%
  \BibitemOpen
  \bibfield  {author} {\bibinfo {author} {\bibfnamefont {K.}~\bibnamefont {Nakazawa}}, \bibinfo {author} {\bibfnamefont {M.}~\bibnamefont {Bibes}},\ and\ \bibinfo {author} {\bibfnamefont {H.}~\bibnamefont {Kohno}},\ }\bibfield  {title} {\bibinfo {title} {Topological {Hall} effect from strong to weak coupling},\ }\href {https://doi.org/10.7566/JPSJ.87.033705} {\bibfield  {journal} {\bibinfo  {journal} {J. Phys. Soc. Jpn.}\ }\textbf {\bibinfo {volume} {87}},\ \bibinfo {pages} {033705} (\bibinfo {year} {2018})}\BibitemShut {NoStop}%
\bibitem [{\citenamefont {Ueda}\ \emph {et~al.}(2012)\citenamefont {Ueda}, \citenamefont {Iguchi}, \citenamefont {Suzuki}, \citenamefont {Ishiwata}, \citenamefont {Taguchi},\ and\ \citenamefont {Tokura}}]{Ueda_2012}%
  \BibitemOpen
  \bibfield  {author} {\bibinfo {author} {\bibfnamefont {K.}~\bibnamefont {Ueda}}, \bibinfo {author} {\bibfnamefont {S.}~\bibnamefont {Iguchi}}, \bibinfo {author} {\bibfnamefont {T.}~\bibnamefont {Suzuki}}, \bibinfo {author} {\bibfnamefont {S.}~\bibnamefont {Ishiwata}}, \bibinfo {author} {\bibfnamefont {Y.}~\bibnamefont {Taguchi}},\ and\ \bibinfo {author} {\bibfnamefont {Y.}~\bibnamefont {Tokura}},\ }\bibfield  {title} {\bibinfo {title} {Topological {Hall} effect in pyrochlore lattice with varying density of spin chirality},\ }\href {https://doi.org/10.1103/PhysRevLett.108.156601} {\bibfield  {journal} {\bibinfo  {journal} {Phys. Rev. Lett.}\ }\textbf {\bibinfo {volume} {108}},\ \bibinfo {pages} {156601} (\bibinfo {year} {2012})}\BibitemShut {NoStop}%
\bibitem [{\citenamefont {Neubauer}\ \emph {et~al.}(2009)\citenamefont {Neubauer}, \citenamefont {Pfleiderer}, \citenamefont {Binz}, \citenamefont {Rosch}, \citenamefont {Ritz}, \citenamefont {Niklowitz},\ and\ \citenamefont {B\"oni}}]{Neubauer_2009}%
  \BibitemOpen
  \bibfield  {author} {\bibinfo {author} {\bibfnamefont {A.}~\bibnamefont {Neubauer}}, \bibinfo {author} {\bibfnamefont {C.}~\bibnamefont {Pfleiderer}}, \bibinfo {author} {\bibfnamefont {B.}~\bibnamefont {Binz}}, \bibinfo {author} {\bibfnamefont {A.}~\bibnamefont {Rosch}}, \bibinfo {author} {\bibfnamefont {R.}~\bibnamefont {Ritz}}, \bibinfo {author} {\bibfnamefont {P.~G.}\ \bibnamefont {Niklowitz}},\ and\ \bibinfo {author} {\bibfnamefont {P.}~\bibnamefont {B\"oni}},\ }\bibfield  {title} {\bibinfo {title} {Topological {Hall} effect in the {$A$} phase of {MnSi}},\ }\href {https://doi.org/10.1103/PhysRevLett.102.186602} {\bibfield  {journal} {\bibinfo  {journal} {Phys. Rev. Lett.}\ }\textbf {\bibinfo {volume} {102}},\ \bibinfo {pages} {186602} (\bibinfo {year} {2009})}\BibitemShut {NoStop}%
\bibitem [{\citenamefont {Metalidis}\ and\ \citenamefont {Bruno}(2006)}]{Metalidis_2006}%
  \BibitemOpen
  \bibfield  {author} {\bibinfo {author} {\bibfnamefont {G.}~\bibnamefont {Metalidis}}\ and\ \bibinfo {author} {\bibfnamefont {P.}~\bibnamefont {Bruno}},\ }\bibfield  {title} {\bibinfo {title} {Topological {Hall} effect studied in simple models},\ }\href {https://doi.org/10.1103/PhysRevB.74.045327} {\bibfield  {journal} {\bibinfo  {journal} {Phys. Rev. B}\ }\textbf {\bibinfo {volume} {74}},\ \bibinfo {pages} {045327} (\bibinfo {year} {2006})}\BibitemShut {NoStop}%
\bibitem [{\citenamefont {Verma}\ \emph {et~al.}(2022)\citenamefont {Verma}, \citenamefont {Addison},\ and\ \citenamefont {Randeria}}]{Verma_2022}%
  \BibitemOpen
  \bibfield  {author} {\bibinfo {author} {\bibfnamefont {N.}~\bibnamefont {Verma}}, \bibinfo {author} {\bibfnamefont {Z.}~\bibnamefont {Addison}},\ and\ \bibinfo {author} {\bibfnamefont {M.}~\bibnamefont {Randeria}},\ }\bibfield  {title} {\bibinfo {title} {Unified theory of the anomalous and topological {Hall} effects with phase-space {Berry} curvatures},\ }\href {https://doi.org/10.1126/sciadv.abq2765} {\bibfield  {journal} {\bibinfo  {journal} {Sci. Adv.}\ }\textbf {\bibinfo {volume} {8}},\ \bibinfo {pages} {eabq2765} (\bibinfo {year} {2022})}\BibitemShut {NoStop}%
\bibitem [{\citenamefont {Go}\ \emph {et~al.}(2025)\citenamefont {Go}, \citenamefont {Goli}, \citenamefont {Esaki}, \citenamefont {Tserkovnyak},\ and\ \citenamefont {Kim}}]{Go_2025}%
  \BibitemOpen
  \bibfield  {author} {\bibinfo {author} {\bibfnamefont {G.}~\bibnamefont {Go}}, \bibinfo {author} {\bibfnamefont {D.~P.}\ \bibnamefont {Goli}}, \bibinfo {author} {\bibfnamefont {N.}~\bibnamefont {Esaki}}, \bibinfo {author} {\bibfnamefont {Y.}~\bibnamefont {Tserkovnyak}},\ and\ \bibinfo {author} {\bibfnamefont {S.~K.}\ \bibnamefont {Kim}},\ }\bibfield  {title} {\bibinfo {title} {Scalar spin chirality {Nernst} effect},\ }\href {https://doi.org/10.1103/PhysRevResearch.7.023190} {\bibfield  {journal} {\bibinfo  {journal} {Phys. Rev. Res.}\ }\textbf {\bibinfo {volume} {7}},\ \bibinfo {pages} {023190} (\bibinfo {year} {2025})}\BibitemShut {NoStop}%
\bibitem [{\citenamefont {Zhang}\ \emph {et~al.}(2021)\citenamefont {Zhang}, \citenamefont {Xu},\ and\ \citenamefont {Ke}}]{Zhang_2021}%
  \BibitemOpen
  \bibfield  {author} {\bibinfo {author} {\bibfnamefont {H.}~\bibnamefont {Zhang}}, \bibinfo {author} {\bibfnamefont {C.~Q.}\ \bibnamefont {Xu}},\ and\ \bibinfo {author} {\bibfnamefont {X.}~\bibnamefont {Ke}},\ }\bibfield  {title} {\bibinfo {title} {Topological {Nernst} effect, anomalous {Nernst} effect, and anomalous thermal {Hall} effect in the {Dirac} semimetal {$\mathrm{Fe}_3\mathrm{Sn}_2$}},\ }\href {https://doi.org/10.1103/PhysRevB.103.L201101} {\bibfield  {journal} {\bibinfo  {journal} {Phys. Rev. B}\ }\textbf {\bibinfo {volume} {103}},\ \bibinfo {pages} {L201101} (\bibinfo {year} {2021})}\BibitemShut {NoStop}%
\bibitem [{\citenamefont {Oike}\ \emph {et~al.}(2022)\citenamefont {Oike}, \citenamefont {Ebino}, \citenamefont {Koretsune}, \citenamefont {Kikkawa}, \citenamefont {Hirschberger}, \citenamefont {Taguchi}, \citenamefont {Tokura},\ and\ \citenamefont {Kagawa}}]{Oike_2022}%
  \BibitemOpen
  \bibfield  {author} {\bibinfo {author} {\bibfnamefont {H.}~\bibnamefont {Oike}}, \bibinfo {author} {\bibfnamefont {T.}~\bibnamefont {Ebino}}, \bibinfo {author} {\bibfnamefont {T.}~\bibnamefont {Koretsune}}, \bibinfo {author} {\bibfnamefont {A.}~\bibnamefont {Kikkawa}}, \bibinfo {author} {\bibfnamefont {M.}~\bibnamefont {Hirschberger}}, \bibinfo {author} {\bibfnamefont {Y.}~\bibnamefont {Taguchi}}, \bibinfo {author} {\bibfnamefont {Y.}~\bibnamefont {Tokura}},\ and\ \bibinfo {author} {\bibfnamefont {F.}~\bibnamefont {Kagawa}},\ }\bibfield  {title} {\bibinfo {title} {Topological {Nernst} effect emerging from real-space gauge field and thermal fluctuations in a magnetic skyrmion lattice},\ }\href {https://doi.org/10.1103/PhysRevB.106.214425} {\bibfield  {journal} {\bibinfo  {journal} {Phys. Rev. B}\ }\textbf {\bibinfo {volume} {106}},\ \bibinfo {pages} {214425} (\bibinfo {year} {2022})}\BibitemShut {NoStop}%
\bibitem [{\citenamefont {Li}\ \emph {et~al.}(2019)\citenamefont {Li}, \citenamefont {Ding}, \citenamefont {Chen}, \citenamefont {Li}, \citenamefont {Hou}, \citenamefont {Liu}, \citenamefont {Zhang}, \citenamefont {Xi}, \citenamefont {Wu},\ and\ \citenamefont {Wang}}]{Li_2019}%
  \BibitemOpen
  \bibfield  {author} {\bibinfo {author} {\bibfnamefont {H.}~\bibnamefont {Li}}, \bibinfo {author} {\bibfnamefont {B.}~\bibnamefont {Ding}}, \bibinfo {author} {\bibfnamefont {J.}~\bibnamefont {Chen}}, \bibinfo {author} {\bibfnamefont {Z.}~\bibnamefont {Li}}, \bibinfo {author} {\bibfnamefont {Z.}~\bibnamefont {Hou}}, \bibinfo {author} {\bibfnamefont {E.}~\bibnamefont {Liu}}, \bibinfo {author} {\bibfnamefont {H.}~\bibnamefont {Zhang}}, \bibinfo {author} {\bibfnamefont {X.}~\bibnamefont {Xi}}, \bibinfo {author} {\bibfnamefont {G.}~\bibnamefont {Wu}},\ and\ \bibinfo {author} {\bibfnamefont {W.}~\bibnamefont {Wang}},\ }\bibfield  {title} {\bibinfo {title} {Large topological {Hall} effect in a geometrically frustrated kagome magnet {Fe$_3$Sn$_2$}},\ }\href {https://doi.org/10.1063/1.5088173} {\bibfield  {journal} {\bibinfo  {journal} {Appl. Phys. Lett.}\ }\textbf {\bibinfo {volume} {114}},\ \bibinfo {pages} {192408} (\bibinfo {year} {2019})}\BibitemShut {NoStop}%
\bibitem [{\citenamefont {Kolincio}\ \emph {et~al.}(2021)\citenamefont {Kolincio}, \citenamefont {Hirschberger}, \citenamefont {Masell}, \citenamefont {Gao}, \citenamefont {Kikkawa}, \citenamefont {Taguchi}, \citenamefont {Arima}, \citenamefont {Nagaosa},\ and\ \citenamefont {Tokura}}]{Kolincio_2021}%
  \BibitemOpen
  \bibfield  {author} {\bibinfo {author} {\bibfnamefont {K.~K.}\ \bibnamefont {Kolincio}}, \bibinfo {author} {\bibfnamefont {M.}~\bibnamefont {Hirschberger}}, \bibinfo {author} {\bibfnamefont {J.}~\bibnamefont {Masell}}, \bibinfo {author} {\bibfnamefont {S.}~\bibnamefont {Gao}}, \bibinfo {author} {\bibfnamefont {A.}~\bibnamefont {Kikkawa}}, \bibinfo {author} {\bibfnamefont {Y.}~\bibnamefont {Taguchi}}, \bibinfo {author} {\bibfnamefont {T.-h.}\ \bibnamefont {Arima}}, \bibinfo {author} {\bibfnamefont {N.}~\bibnamefont {Nagaosa}},\ and\ \bibinfo {author} {\bibfnamefont {Y.}~\bibnamefont {Tokura}},\ }\bibfield  {title} {\bibinfo {title} {Large {Hall} and {Nernst} responses from thermally induced spin chirality in a spin-trimer ferromagnet},\ }\href {https://doi.org/10.1073/pnas.2023588118} {\bibfield  {journal} {\bibinfo  {journal} {Proc. Natl. Acad. Sci. U.S.A.}\ }\textbf {\bibinfo {volume} {118}},\ \bibinfo {pages} {e2023588118} (\bibinfo {year} {2021})}\BibitemShut {NoStop}%
\bibitem [{\citenamefont {Vistoli}\ \emph {et~al.}(2018)\citenamefont {Vistoli}, \citenamefont {Wang}, \citenamefont {Sander}, \citenamefont {Zhu}, \citenamefont {Casals}, \citenamefont {Cichelero}, \citenamefont {Barth{\'e}l{\'e}my}, \citenamefont {Fusil}, \citenamefont {Herranz}, \citenamefont {Valencia}, \citenamefont {Abrudan}, \citenamefont {Weschke}, \citenamefont {Nakazawa}, \citenamefont {Kohno}, \citenamefont {Santamaria}, \citenamefont {Wu}, \citenamefont {Garcia},\ and\ \citenamefont {Bibes}}]{Vistoli_2018}%
  \BibitemOpen
  \bibfield  {author} {\bibinfo {author} {\bibfnamefont {L.}~\bibnamefont {Vistoli}}, \bibinfo {author} {\bibfnamefont {W.}~\bibnamefont {Wang}}, \bibinfo {author} {\bibfnamefont {A.}~\bibnamefont {Sander}}, \bibinfo {author} {\bibfnamefont {Q.}~\bibnamefont {Zhu}}, \bibinfo {author} {\bibfnamefont {B.}~\bibnamefont {Casals}}, \bibinfo {author} {\bibfnamefont {R.}~\bibnamefont {Cichelero}}, \bibinfo {author} {\bibfnamefont {A.}~\bibnamefont {Barth{\'e}l{\'e}my}}, \bibinfo {author} {\bibfnamefont {S.}~\bibnamefont {Fusil}}, \bibinfo {author} {\bibfnamefont {G.}~\bibnamefont {Herranz}}, \bibinfo {author} {\bibfnamefont {S.}~\bibnamefont {Valencia}}, \bibinfo {author} {\bibfnamefont {R.}~\bibnamefont {Abrudan}}, \bibinfo {author} {\bibfnamefont {E.}~\bibnamefont {Weschke}}, \bibinfo {author} {\bibfnamefont {K.}~\bibnamefont {Nakazawa}}, \bibinfo {author} {\bibfnamefont {H.}~\bibnamefont {Kohno}}, \bibinfo {author} {\bibfnamefont {J.}~\bibnamefont {Santamaria}}, \bibinfo {author} {\bibfnamefont {W.}~\bibnamefont {Wu}}, \bibinfo {author} {\bibfnamefont {V.}~\bibnamefont {Garcia}},\ and\ \bibinfo {author} {\bibfnamefont {M.}~\bibnamefont {Bibes}},\ }\bibfield  {title} {\bibinfo {title} {Giant topological {Hall} effect in correlated oxide thin films},\ }\href {https://doi.org/10.1038/s41567-018-0307-5} {\bibfield  {journal} {\bibinfo  {journal} {Nat. Phys.}\ }\textbf {\bibinfo {volume} {15}},\ \bibinfo {pages} {67} (\bibinfo {year} {2018})}\BibitemShut {NoStop}%
\bibitem [{\citenamefont {Nakazawa}\ and\ \citenamefont {Kohno}(2014)}]{Nakazawa_2014}%
  \BibitemOpen
  \bibfield  {author} {\bibinfo {author} {\bibfnamefont {K.}~\bibnamefont {Nakazawa}}\ and\ \bibinfo {author} {\bibfnamefont {H.}~\bibnamefont {Kohno}},\ }\bibfield  {title} {\bibinfo {title} {Effects of vertex corrections on the chirality-driven anomalous {Hall} effect},\ }\href {https://doi.org/10.7566/JPSJ.83.073707} {\bibfield  {journal} {\bibinfo  {journal} {J. Phys. Soc. Jpn.}\ }\textbf {\bibinfo {volume} {83}},\ \bibinfo {pages} {073707} (\bibinfo {year} {2014})}\BibitemShut {NoStop}%
\bibitem [{\citenamefont {Onoda}\ \emph {et~al.}(2004)\citenamefont {Onoda}, \citenamefont {Tatara},\ and\ \citenamefont {Nagaosa}}]{Onoda_2004}%
  \BibitemOpen
  \bibfield  {author} {\bibinfo {author} {\bibfnamefont {M.}~\bibnamefont {Onoda}}, \bibinfo {author} {\bibfnamefont {G.}~\bibnamefont {Tatara}},\ and\ \bibinfo {author} {\bibfnamefont {N.}~\bibnamefont {Nagaosa}},\ }\bibfield  {title} {\bibinfo {title} {Anomalous {Hall} effect and skyrmion number in real and momentum spaces},\ }\href {https://doi.org/10.1143/JPSJ.73.2624} {\bibfield  {journal} {\bibinfo  {journal} {J. Phys. Soc. Jpn.}\ }\textbf {\bibinfo {volume} {73}},\ \bibinfo {pages} {2624} (\bibinfo {year} {2004})}\BibitemShut {NoStop}%
\bibitem [{\citenamefont {Vir}\ \emph {et~al.}(2019)\citenamefont {Vir}, \citenamefont {Gayles}, \citenamefont {Sukhanov}, \citenamefont {Kumar}, \citenamefont {Damay}, \citenamefont {Sun}, \citenamefont {K\"ubler}, \citenamefont {Shekhar},\ and\ \citenamefont {Felser}}]{Vir_2019}%
  \BibitemOpen
  \bibfield  {author} {\bibinfo {author} {\bibfnamefont {P.}~\bibnamefont {Vir}}, \bibinfo {author} {\bibfnamefont {J.}~\bibnamefont {Gayles}}, \bibinfo {author} {\bibfnamefont {A.~S.}\ \bibnamefont {Sukhanov}}, \bibinfo {author} {\bibfnamefont {N.}~\bibnamefont {Kumar}}, \bibinfo {author} {\bibfnamefont {F.~m.~c.}\ \bibnamefont {Damay}}, \bibinfo {author} {\bibfnamefont {Y.}~\bibnamefont {Sun}}, \bibinfo {author} {\bibfnamefont {J.}~\bibnamefont {K\"ubler}}, \bibinfo {author} {\bibfnamefont {C.}~\bibnamefont {Shekhar}},\ and\ \bibinfo {author} {\bibfnamefont {C.}~\bibnamefont {Felser}},\ }\bibfield  {title} {\bibinfo {title} {Anisotropic topological {Hall} effect with real and momentum space {Berry} curvature in the antiskrymion-hosting heusler compound {Mn$_{1.4}$PtSn}},\ }\href {https://doi.org/10.1103/PhysRevB.99.140406} {\bibfield  {journal} {\bibinfo  {journal} {Phys. Rev. B}\ }\textbf {\bibinfo {volume} {99}},\ \bibinfo {pages} {140406} (\bibinfo {year} {2019})}\BibitemShut {NoStop}%
\bibitem [{\citenamefont {Feng}\ \emph {et~al.}(2020)\citenamefont {Feng}, \citenamefont {Hanke}, \citenamefont {Zhou}, \citenamefont {Guo}, \citenamefont {Bl{\"u}gel}, \citenamefont {Mokrousov},\ and\ \citenamefont {Yao}}]{Feng_2020}%
  \BibitemOpen
  \bibfield  {author} {\bibinfo {author} {\bibfnamefont {W.}~\bibnamefont {Feng}}, \bibinfo {author} {\bibfnamefont {J.-P.}\ \bibnamefont {Hanke}}, \bibinfo {author} {\bibfnamefont {X.}~\bibnamefont {Zhou}}, \bibinfo {author} {\bibfnamefont {G.-Y.}\ \bibnamefont {Guo}}, \bibinfo {author} {\bibfnamefont {S.}~\bibnamefont {Bl{\"u}gel}}, \bibinfo {author} {\bibfnamefont {Y.}~\bibnamefont {Mokrousov}},\ and\ \bibinfo {author} {\bibfnamefont {Y.}~\bibnamefont {Yao}},\ }\bibfield  {title} {\bibinfo {title} {Topological magneto-optical effects and their quantization in noncoplanar antiferromagnets},\ }\href {https://doi.org/10.1038/s41467-019-13968-8} {\bibfield  {journal} {\bibinfo  {journal} {Nat. Commun.}\ }\textbf {\bibinfo {volume} {11}},\ \bibinfo {pages} {118} (\bibinfo {year} {2020})}\BibitemShut {NoStop}%
\bibitem [{\citenamefont {Xiao}\ \emph {et~al.}(2006)\citenamefont {Xiao}, \citenamefont {Yao}, \citenamefont {Fang},\ and\ \citenamefont {Niu}}]{Xiao_2006_PRL}%
  \BibitemOpen
  \bibfield  {author} {\bibinfo {author} {\bibfnamefont {D.}~\bibnamefont {Xiao}}, \bibinfo {author} {\bibfnamefont {Y.}~\bibnamefont {Yao}}, \bibinfo {author} {\bibfnamefont {Z.}~\bibnamefont {Fang}},\ and\ \bibinfo {author} {\bibfnamefont {Q.}~\bibnamefont {Niu}},\ }\bibfield  {title} {\bibinfo {title} {{Berry}-phase effect in anomalous thermoelectric transport},\ }\href {https://doi.org/10.1103/PhysRevLett.97.026603} {\bibfield  {journal} {\bibinfo  {journal} {Phys. Rev. Lett.}\ }\textbf {\bibinfo {volume} {97}},\ \bibinfo {pages} {026603} (\bibinfo {year} {2006})}\BibitemShut {NoStop}%
\bibitem [{\citenamefont {Ishizuka}\ and\ \citenamefont {Nagaosa}(2018)}]{Ishizuka_2018}%
  \BibitemOpen
  \bibfield  {author} {\bibinfo {author} {\bibfnamefont {H.}~\bibnamefont {Ishizuka}}\ and\ \bibinfo {author} {\bibfnamefont {N.}~\bibnamefont {Nagaosa}},\ }\bibfield  {title} {\bibinfo {title} {Spin chirality induced skew scattering and anomalous {Hall} effect in chiral magnets},\ }\href {https://doi.org/10.1126/sciadv.aap9962} {\bibfield  {journal} {\bibinfo  {journal} {Sci. Adv.}\ }\textbf {\bibinfo {volume} {4}},\ \bibinfo {pages} {eaap9962} (\bibinfo {year} {2018})}\BibitemShut {NoStop}%
\bibitem [{\citenamefont {Chen}\ \emph {et~al.}(2014)\citenamefont {Chen}, \citenamefont {Niu},\ and\ \citenamefont {MacDonald}}]{Chen_2014}%
  \BibitemOpen
  \bibfield  {author} {\bibinfo {author} {\bibfnamefont {H.}~\bibnamefont {Chen}}, \bibinfo {author} {\bibfnamefont {Q.}~\bibnamefont {Niu}},\ and\ \bibinfo {author} {\bibfnamefont {A.~H.}\ \bibnamefont {MacDonald}},\ }\bibfield  {title} {\bibinfo {title} {Anomalous {H}all effect arising from noncollinear antiferromagnetism},\ }\href {https://doi.org/10.1103/PhysRevLett.112.017205} {\bibfield  {journal} {\bibinfo  {journal} {Phys. Rev. Lett.}\ }\textbf {\bibinfo {volume} {112}},\ \bibinfo {pages} {017205} (\bibinfo {year} {2014})}\BibitemShut {NoStop}%
\bibitem [{\citenamefont {Syromyatnikov}\ and\ \citenamefont {Maleyev}(2002)}]{Syromiatnikov_2002}%
  \BibitemOpen
  \bibfield  {author} {\bibinfo {author} {\bibfnamefont {A.~V.}\ \bibnamefont {Syromyatnikov}}\ and\ \bibinfo {author} {\bibfnamefont {S.~V.}\ \bibnamefont {Maleyev}},\ }\bibfield  {title} {\bibinfo {title} {Hidden long-range order in kagom\'e {H}eisenberg antiferromagnets},\ }\href {https://doi.org/10.1103/PhysRevB.66.132408} {\bibfield  {journal} {\bibinfo  {journal} {Phys. Rev. B}\ }\textbf {\bibinfo {volume} {66}},\ \bibinfo {pages} {132408} (\bibinfo {year} {2002})}\BibitemShut {NoStop}%
\bibitem [{\citenamefont {Han}(2017)}]{Han_2017}%
  \BibitemOpen
  \bibfield  {author} {\bibinfo {author} {\bibfnamefont {J.~H.}\ \bibnamefont {Han}},\ }\href {https://doi.org/10.1007/978-3-319-69246-3} {\emph {\bibinfo {title} {{S}kyrmions in Condensed Matter}}},\ \bibinfo {series} {Springer Tracts in Modern Physics}, Vol.\ \bibinfo {volume} {278}\ (\bibinfo  {publisher} {Springer International Publishing},\ \bibinfo {address} {Cham},\ \bibinfo {year} {2017})\BibitemShut {NoStop}%
\bibitem [{\citenamefont {Ohgushi}\ \emph {et~al.}(2000)\citenamefont {Ohgushi}, \citenamefont {Murakami},\ and\ \citenamefont {Nagaosa}}]{Ohgushi_2000}%
  \BibitemOpen
  \bibfield  {author} {\bibinfo {author} {\bibfnamefont {K.}~\bibnamefont {Ohgushi}}, \bibinfo {author} {\bibfnamefont {S.}~\bibnamefont {Murakami}},\ and\ \bibinfo {author} {\bibfnamefont {N.}~\bibnamefont {Nagaosa}},\ }\bibfield  {title} {\bibinfo {title} {Spin chirality and anomalous hall effect in itinerant ferromagnets},\ }\href {https://doi.org/10.1103/PhysRevB.62.R6065} {\bibfield  {journal} {\bibinfo  {journal} {Phys. Rev. B}\ }\textbf {\bibinfo {volume} {62}},\ \bibinfo {pages} {R6065} (\bibinfo {year} {2000})}\BibitemShut {NoStop}%
\bibitem [{\citenamefont {Shindou}\ and\ \citenamefont {Nagaosa}(2001)}]{Shindou_2001}%
  \BibitemOpen
  \bibfield  {author} {\bibinfo {author} {\bibfnamefont {R.}~\bibnamefont {Shindou}}\ and\ \bibinfo {author} {\bibfnamefont {N.}~\bibnamefont {Nagaosa}},\ }\bibfield  {title} {\bibinfo {title} {Orbital ferromagnetism and anomalous hall effect in antiferromagnets on the kagome lattice},\ }\href {https://doi.org/10.1103/PhysRevLett.87.116801} {\bibfield  {journal} {\bibinfo  {journal} {Phys. Rev. Lett.}\ }\textbf {\bibinfo {volume} {87}},\ \bibinfo {pages} {116801} (\bibinfo {year} {2001})}\BibitemShut {NoStop}%
\bibitem [{\citenamefont {Yang}\ \emph {et~al.}(2015)\citenamefont {Yang}, \citenamefont {Thiaville}, \citenamefont {Rohart}, \citenamefont {Fert},\ and\ \citenamefont {Chshiev}}]{Yang_2015}%
  \BibitemOpen
  \bibfield  {author} {\bibinfo {author} {\bibfnamefont {H.}~\bibnamefont {Yang}}, \bibinfo {author} {\bibfnamefont {A.}~\bibnamefont {Thiaville}}, \bibinfo {author} {\bibfnamefont {S.}~\bibnamefont {Rohart}}, \bibinfo {author} {\bibfnamefont {A.}~\bibnamefont {Fert}},\ and\ \bibinfo {author} {\bibfnamefont {M.}~\bibnamefont {Chshiev}},\ }\bibfield  {title} {\bibinfo {title} {Anatomy of {D}zyaloshinskii-{M}oriya interaction at {Co}/{Pt} interfaces},\ }\href {https://doi.org/10.1103/PhysRevLett.115.267210} {\bibfield  {journal} {\bibinfo  {journal} {Phys. Rev. Lett.}\ }\textbf {\bibinfo {volume} {115}},\ \bibinfo {pages} {267210} (\bibinfo {year} {2015})}\BibitemShut {NoStop}%
\bibitem [{\citenamefont {Liang}\ \emph {et~al.}(2020)\citenamefont {Liang}, \citenamefont {Wang}, \citenamefont {Du}, \citenamefont {Hallal}, \citenamefont {Garcia}, \citenamefont {Chshiev}, \citenamefont {Fert},\ and\ \citenamefont {Yang}}]{Liang_2020}%
  \BibitemOpen
  \bibfield  {author} {\bibinfo {author} {\bibfnamefont {J.}~\bibnamefont {Liang}}, \bibinfo {author} {\bibfnamefont {W.}~\bibnamefont {Wang}}, \bibinfo {author} {\bibfnamefont {H.}~\bibnamefont {Du}}, \bibinfo {author} {\bibfnamefont {A.}~\bibnamefont {Hallal}}, \bibinfo {author} {\bibfnamefont {K.}~\bibnamefont {Garcia}}, \bibinfo {author} {\bibfnamefont {M.}~\bibnamefont {Chshiev}}, \bibinfo {author} {\bibfnamefont {A.}~\bibnamefont {Fert}},\ and\ \bibinfo {author} {\bibfnamefont {H.}~\bibnamefont {Yang}},\ }\bibfield  {title} {\bibinfo {title} {Very large {D}zyaloshinskii-{M}oriya interaction in two-dimensional {J}anus manganese dichalcogenides and its application to realize skyrmion states},\ }\href {https://doi.org/10.1103/PhysRevB.101.184401} {\bibfield  {journal} {\bibinfo  {journal} {Phys. Rev. B}\ }\textbf {\bibinfo {volume} {101}},\ \bibinfo {pages} {184401} (\bibinfo {year} {2020})}\BibitemShut {NoStop}%
\bibitem [{\citenamefont {Koshibae}\ and\ \citenamefont {Nagaosa}(2016)}]{Koshibae_2016}%
  \BibitemOpen
  \bibfield  {author} {\bibinfo {author} {\bibfnamefont {W.}~\bibnamefont {Koshibae}}\ and\ \bibinfo {author} {\bibfnamefont {N.}~\bibnamefont {Nagaosa}},\ }\bibfield  {title} {\bibinfo {title} {Theory of antiskyrmions in magnets},\ }\href {https://doi.org/10.1038/ncomms10542} {\bibfield  {journal} {\bibinfo  {journal} {Nat. Commun.}\ }\textbf {\bibinfo {volume} {7}},\ \bibinfo {pages} {10542} (\bibinfo {year} {2016})}\BibitemShut {NoStop}%
\bibitem [{\citenamefont {H\"artl}\ \emph {et~al.}(2024)\citenamefont {H\"artl}, \citenamefont {Vogt}, \citenamefont {Leisegang}, \citenamefont {Bihlmayer}, \citenamefont {Bl\"ugel},\ and\ \citenamefont {Bode}}]{Hartl_2024}%
  \BibitemOpen
  \bibfield  {author} {\bibinfo {author} {\bibfnamefont {P.}~\bibnamefont {H\"artl}}, \bibinfo {author} {\bibfnamefont {M.}~\bibnamefont {Vogt}}, \bibinfo {author} {\bibfnamefont {M.}~\bibnamefont {Leisegang}}, \bibinfo {author} {\bibfnamefont {G.}~\bibnamefont {Bihlmayer}}, \bibinfo {author} {\bibfnamefont {S.}~\bibnamefont {Bl\"ugel}},\ and\ \bibinfo {author} {\bibfnamefont {M.}~\bibnamefont {Bode}},\ }\bibfield  {title} {\bibinfo {title} {Spin spiral state at a ferromagnetic {G}d vacuum interface},\ }\href {https://doi.org/10.1103/PhysRevLett.133.186701} {\bibfield  {journal} {\bibinfo  {journal} {Phys. Rev. Lett.}\ }\textbf {\bibinfo {volume} {133}},\ \bibinfo {pages} {186701} (\bibinfo {year} {2024})}\BibitemShut {NoStop}%
\bibitem [{\citenamefont {Rosales}\ \emph {et~al.}(2023)\citenamefont {Rosales}, \citenamefont {G\'omez~Albarrac\'{\i}n}, \citenamefont {Pujol},\ and\ \citenamefont {Jaubert}}]{Rosales_2023}%
  \BibitemOpen
  \bibfield  {author} {\bibinfo {author} {\bibfnamefont {H.~D.}\ \bibnamefont {Rosales}}, \bibinfo {author} {\bibfnamefont {F.~A.}\ \bibnamefont {G\'omez~Albarrac\'{\i}n}}, \bibinfo {author} {\bibfnamefont {P.}~\bibnamefont {Pujol}},\ and\ \bibinfo {author} {\bibfnamefont {L.~D.~C.}\ \bibnamefont {Jaubert}},\ }\bibfield  {title} {\bibinfo {title} {Skyrmion fluid and bimeron glass protected by a chiral spin liquid on a {K}agome lattice},\ }\href {https://doi.org/10.1103/PhysRevLett.130.106703} {\bibfield  {journal} {\bibinfo  {journal} {Phys. Rev. Lett.}\ }\textbf {\bibinfo {volume} {130}},\ \bibinfo {pages} {106703} (\bibinfo {year} {2023})}\BibitemShut {NoStop}%
\bibitem [{\citenamefont {G\'omez~Albarrac\'{\i}n}(2024)}]{Albarracin_2024_ML}%
  \BibitemOpen
  \bibfield  {author} {\bibinfo {author} {\bibfnamefont {F.~A.}\ \bibnamefont {G\'omez~Albarrac\'{\i}n}},\ }\bibfield  {title} {\bibinfo {title} {Unsupervised machine learning for the detection of exotic phases in skyrmion phase diagrams},\ }\href {https://doi.org/10.1103/PhysRevB.110.214415} {\bibfield  {journal} {\bibinfo  {journal} {Phys. Rev. B}\ }\textbf {\bibinfo {volume} {110}},\ \bibinfo {pages} {214415} (\bibinfo {year} {2024})}\BibitemShut {NoStop}%
\bibitem [{\citenamefont {G\'omez~Albarrac\'{\i}n}\ \emph {et~al.}(2024)\citenamefont {G\'omez~Albarrac\'{\i}n}, \citenamefont {Rosales}, \citenamefont {Udagawa}, \citenamefont {Pujol},\ and\ \citenamefont {Jaubert}}]{Albarracin_2024_PRB}%
  \BibitemOpen
  \bibfield  {author} {\bibinfo {author} {\bibfnamefont {F.~A.}\ \bibnamefont {G\'omez~Albarrac\'{\i}n}}, \bibinfo {author} {\bibfnamefont {H.~D.}\ \bibnamefont {Rosales}}, \bibinfo {author} {\bibfnamefont {M.}~\bibnamefont {Udagawa}}, \bibinfo {author} {\bibfnamefont {P.}~\bibnamefont {Pujol}},\ and\ \bibinfo {author} {\bibfnamefont {L.~D.~C.}\ \bibnamefont {Jaubert}},\ }\bibfield  {title} {\bibinfo {title} {From chiral spin liquids to skyrmion fluids and crystals, and their interplay with itinerant electrons},\ }\href {https://doi.org/10.1103/PhysRevB.109.064426} {\bibfield  {journal} {\bibinfo  {journal} {Phys. Rev. B}\ }\textbf {\bibinfo {volume} {109}},\ \bibinfo {pages} {064426} (\bibinfo {year} {2024})}\BibitemShut {NoStop}%
\bibitem [{\citenamefont {Kim}\ and\ \citenamefont {Mulkers}(2020)}]{Kim_2020}%
  \BibitemOpen
  \bibfield  {author} {\bibinfo {author} {\bibfnamefont {J.-V.}\ \bibnamefont {Kim}}\ and\ \bibinfo {author} {\bibfnamefont {J.}~\bibnamefont {Mulkers}},\ }\bibfield  {title} {\bibinfo {title} {On quantifying the topological charge in micromagnetics using a lattice-based approach},\ }\href {https://doi.org/10.1088/2633-1357/abad0c} {\bibfield  {journal} {\bibinfo  {journal} {IOP SciNotes}\ }\textbf {\bibinfo {volume} {1}},\ \bibinfo {pages} {025211} (\bibinfo {year} {2020})}\BibitemShut {NoStop}%
\bibitem [{\citenamefont {Nagaosa}\ and\ \citenamefont {Tokura}(2013)}]{Nagaosa_2013}%
  \BibitemOpen
  \bibfield  {author} {\bibinfo {author} {\bibfnamefont {N.}~\bibnamefont {Nagaosa}}\ and\ \bibinfo {author} {\bibfnamefont {Y.}~\bibnamefont {Tokura}},\ }\bibfield  {title} {\bibinfo {title} {Topological properties and dynamics of magnetic skyrmions},\ }\href {https://doi.org/10.1038/nnano.2013.243} {\bibfield  {journal} {\bibinfo  {journal} {Nature Nanotechnology}\ }\textbf {\bibinfo {volume} {8}},\ \bibinfo {pages} {899} (\bibinfo {year} {2013})}\BibitemShut {NoStop}%
\bibitem [{\citenamefont {Xie}\ \emph {et~al.}(2021)\citenamefont {Xie},
  \citenamefont {Chen}, \citenamefont {Chen}, \citenamefont {Wang},
  \citenamefont {Yin}, \citenamefont {Stewart}, \citenamefont {Stone},
  \citenamefont {Daemen}, \citenamefont {Feng}, \citenamefont {Cao},
  \citenamefont {Lei}, \citenamefont {Yin}, \citenamefont {MacDonald},\ and\
  \citenamefont {Dai}}]{Xie_Chen_2021}%
  \BibitemOpen
  \bibfield  {author} {\bibinfo {author} {\bibfnamefont {Y.}~\bibnamefont
  {Xie}}, \bibinfo {author} {\bibfnamefont {L.}~\bibnamefont {Chen}}, \bibinfo
  {author} {\bibfnamefont {T.}~\bibnamefont {Chen}}, \bibinfo {author}
  {\bibfnamefont {Q.}~\bibnamefont {Wang}}, \bibinfo {author} {\bibfnamefont
  {Q.}~\bibnamefont {Yin}}, \bibinfo {author} {\bibfnamefont {J.~R.}\
  \bibnamefont {Stewart}}, \bibinfo {author} {\bibfnamefont {M.~B.}\
  \bibnamefont {Stone}}, \bibinfo {author} {\bibfnamefont {L.~L.}\ \bibnamefont
  {Daemen}}, \bibinfo {author} {\bibfnamefont {E.}~\bibnamefont {Feng}},
  \bibinfo {author} {\bibfnamefont {H.}~\bibnamefont {Cao}}, \bibinfo {author}
  {\bibfnamefont {H.}~\bibnamefont {Lei}}, \bibinfo {author} {\bibfnamefont
  {Z.}~\bibnamefont {Yin}}, \bibinfo {author} {\bibfnamefont {A.~H.}\
  \bibnamefont {MacDonald}},\ and\ \bibinfo {author} {\bibfnamefont
  {P.}~\bibnamefont {Dai}},\ }\bibfield  {title} {\bibinfo {title} {Spin
  excitations in metallic kagome lattice {FeSn} and {CoSn}},\ }\href
  {https://doi.org/10.1038/s42005-021-00736-8} {\bibfield  {journal} {\bibinfo
  {journal} {Communications Physics}\ }\textbf {\bibinfo {volume} {4}},\
  \bibinfo {pages} {10.1038/s42005-021-00736-8} (\bibinfo {year}
  {2021})}\BibitemShut {NoStop}%
\bibitem [{\citenamefont {Fang}\ \emph {et~al.}(2022)\citenamefont {Fang},
  \citenamefont {Ye}, \citenamefont {Ghimire}, \citenamefont {Kang},
  \citenamefont {Liu}, \citenamefont {Han}, \citenamefont {Fu}, \citenamefont
  {Richter}, \citenamefont {van~den Brink}, \citenamefont {Kaxiras},
  \citenamefont {Comin},\ and\ \citenamefont
  {Checkelsky}}]{PhysRevB.105.035107}%
  \BibitemOpen
  \bibfield  {author} {\bibinfo {author} {\bibfnamefont {S.}~\bibnamefont
  {Fang}}, \bibinfo {author} {\bibfnamefont {L.}~\bibnamefont {Ye}}, \bibinfo
  {author} {\bibfnamefont {M.~P.}\ \bibnamefont {Ghimire}}, \bibinfo {author}
  {\bibfnamefont {M.}~\bibnamefont {Kang}}, \bibinfo {author} {\bibfnamefont
  {J.}~\bibnamefont {Liu}}, \bibinfo {author} {\bibfnamefont {M.}~\bibnamefont
  {Han}}, \bibinfo {author} {\bibfnamefont {L.}~\bibnamefont {Fu}}, \bibinfo
  {author} {\bibfnamefont {M.}~\bibnamefont {Richter}}, \bibinfo {author}
  {\bibfnamefont {J.}~\bibnamefont {van~den Brink}}, \bibinfo {author}
  {\bibfnamefont {E.}~\bibnamefont {Kaxiras}}, \bibinfo {author} {\bibfnamefont
  {R.}~\bibnamefont {Comin}},\ and\ \bibinfo {author} {\bibfnamefont {J.~G.}\
  \bibnamefont {Checkelsky}},\ }\bibfield  {title} {\bibinfo {title}
  {Ferromagnetic helical nodal line and {Kane-Mele} spin-orbit coupling in
  kagome metal {${\mathrm{Fe}}_{3}{\mathrm{Sn}}_{2}$}},\ }\href
  {https://doi.org/10.1103/PhysRevB.105.035107} {\bibfield  {journal} {\bibinfo
   {journal} {Phys. Rev. B}\ }\textbf {\bibinfo {volume} {105}},\ \bibinfo
  {pages} {035107} (\bibinfo {year} {2022})}\BibitemShut {NoStop}%
\bibitem [{\citenamefont {Barros}\ \emph {et~al.}(2014)\citenamefont {Barros}, \citenamefont {Venderbos}, \citenamefont {Chern},\ and\ \citenamefont {Batista}}]{Jsd0.2t}%
  \BibitemOpen
  \bibfield  {author} {\bibinfo {author} {\bibfnamefont {K.}~\bibnamefont {Barros}},
  \bibinfo {author} {\bibfnamefont {J.~W.~F.}\ \bibnamefont {Venderbos}},
  \bibinfo {author} {\bibfnamefont {G.-W.}\ \bibnamefont {Chern}},
  \ and\ \bibinfo {author} {\bibfnamefont {C.~D.}\ \bibnamefont {Batista}},\ }
  \bibfield  {title} {\bibinfo {title} {Exotic magnetic orderings in the kagome Kondo-lattice model},\ }
  \href {https://doi.org/10.1103/PhysRevB.90.245119}
  {\bibfield  {journal} {\bibinfo  {journal} {Phys. Rev. B}\ }
  \textbf {\bibinfo {volume} {90}},\ \bibinfo {pages} {245119}
  (\bibinfo {year} {2014})}\BibitemShut {NoStop}%
\end{thebibliography}
%

\end{document}